\documentclass[10pt,twocolumn,letterpaper]{article}

\usepackage{wacv}
\usepackage{times}
\usepackage{epsfig}
\usepackage{graphicx}
\usepackage{amsmath}
\usepackage{amssymb}
\usepackage{adjustbox}

\usepackage{siunitx}
\usepackage{mathrsfs}
\usepackage{booktabs}
\usepackage{multirow}
\usepackage{stfloats}
\usepackage[caption=false]{subfig}

\usepackage{array}
\newcolumntype{C}[1]{>{\centering\let\newline\\\arraybackslash\hspace{0pt}}m{#1}}

\usepackage[pagebackref=true,breaklinks=true,letterpaper=true,colorlinks,bookmarks=false]{hyperref}

\wacvfinalcopy 

\def\wacvPaperID{} 

\ifwacvfinal\fi
\begin{document}

\title{Unsupervised Learning for Real-World Super-Resolution}

\newcommand{\aand}{\hspace{6mm}}
\author{Andreas Lugmayr \aand  Martin Danelljan \aand Radu Timofte \vspace{1.5mm}\\
	CVL, ETH Z\"urich, Switzerland
}

\maketitle

\begin{abstract}

	Most current super-resolution methods rely on low and high resolution image pairs to train a network in a fully supervised manner. However, such image pairs are not available in real-world applications. Instead of directly addressing this problem, most works employ the popular bicubic downsampling strategy to artificially generate a corresponding low resolution image. Unfortunately, this strategy introduces significant artifacts, removing natural sensor noise and other real-world characteristics. Super-resolution networks trained on such bicubic images therefore struggle to generalize to natural images.
	
	In this work, we propose an unsupervised approach for image super-resolution. Given only unpaired data, we learn to invert the effects of bicubic downsampling in order to restore the natural image characteristics present in the data. This allows us to generate realistic image pairs, faithfully reflecting the distribution of real-world images. Our super-resolution network can therefore be trained with direct pixel-wise supervision in the high resolution domain, while robustly generalizing to real input. We demonstrate the effectiveness of our approach in quantitative and qualitative experiments.
	
\end{abstract}

\section{Introduction}

Super-resolution (SR) aims to enhance the resolution of \emph{natural} images. Recent years have seen an increased interest in the problem, driven by emerging applications. Most notably, current generations of smartphones allow for the deployment of powerful image enhancement techniques, based on machine learning approaches. This calls for super-resolution methods that can be applied to natural images, that are often subject to significant levels of sensor noise, compression artifacts or other corruptions encountered in applications. In this work, we therefore address the problem of super-resolution in the real-world setting.

Real-world SR poses a fundamental challenge that has been largely ignored until very recently. The lack of \emph{natural} low resolution (LR) and high resolution (HR) image pairs greatly complicates the evaluation and training of SR methods. Therefore, research in the field has long relied on the use of known degradation operators such as bicubic kernel in order to artificially generate a corresponding LR image~\cite{dong2014learning, tai2017memnet, tong2017image}. While this straight-forward approach enables simple and efficient benchmarking and generation of virtually unlimited training data, it comes with significant drawbacks. Bicubic downsampling can drastically change the natural characteristics of an image by, \eg, removing sensor noise and compression artifacts.

\begin{figure}
\centering

\includegraphics[width=0.24\linewidth]{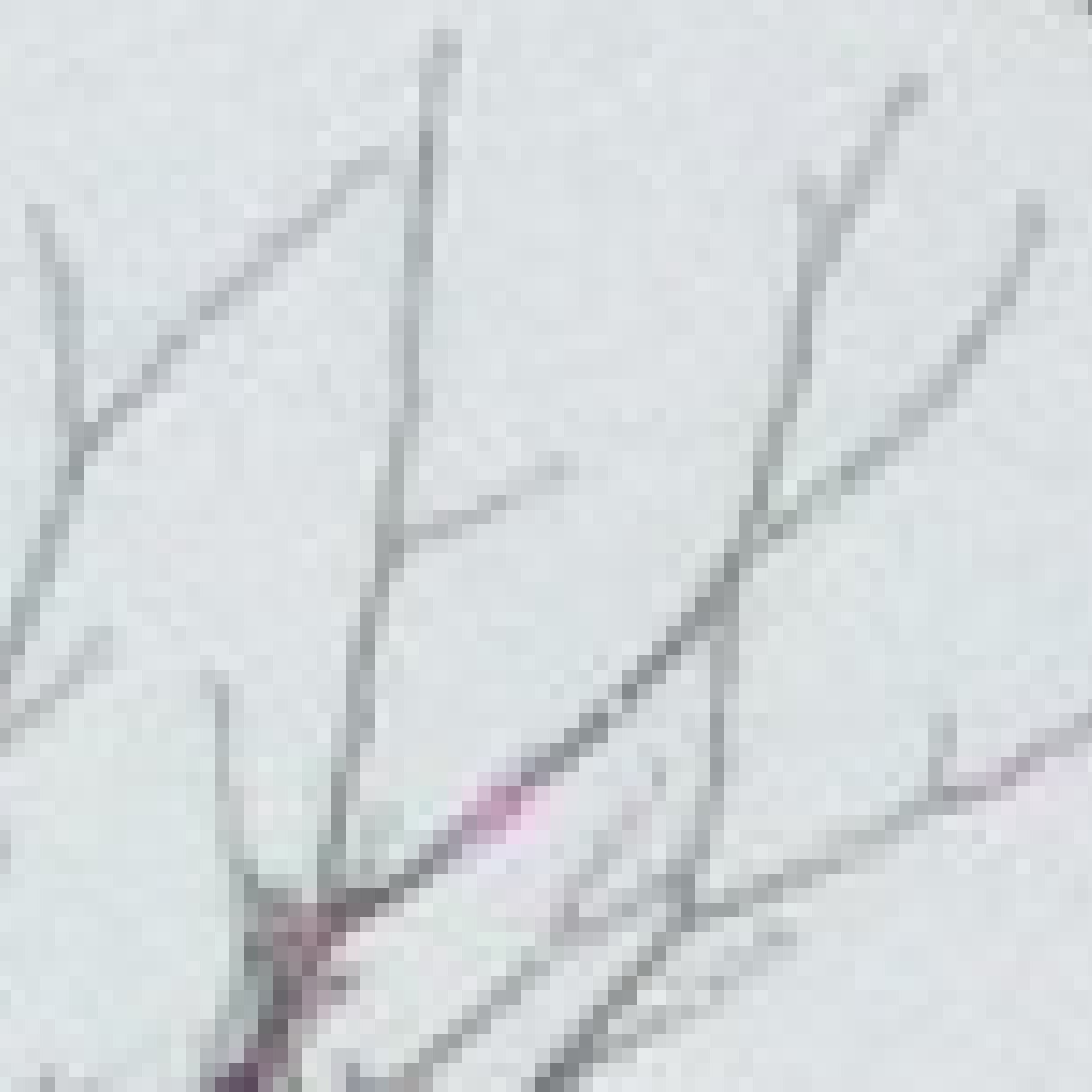}
\includegraphics[width=0.24\linewidth]{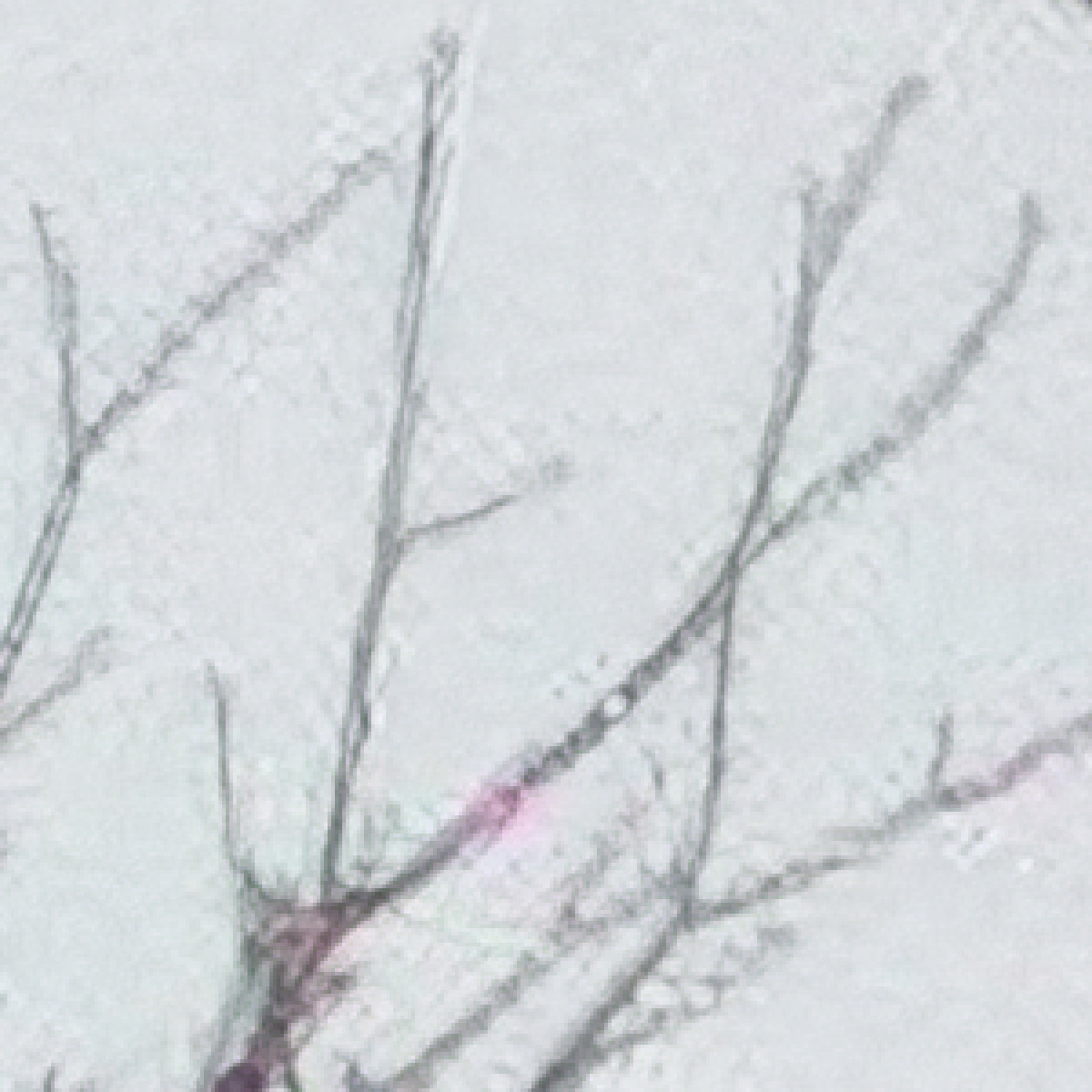}
\includegraphics[width=0.24\linewidth]{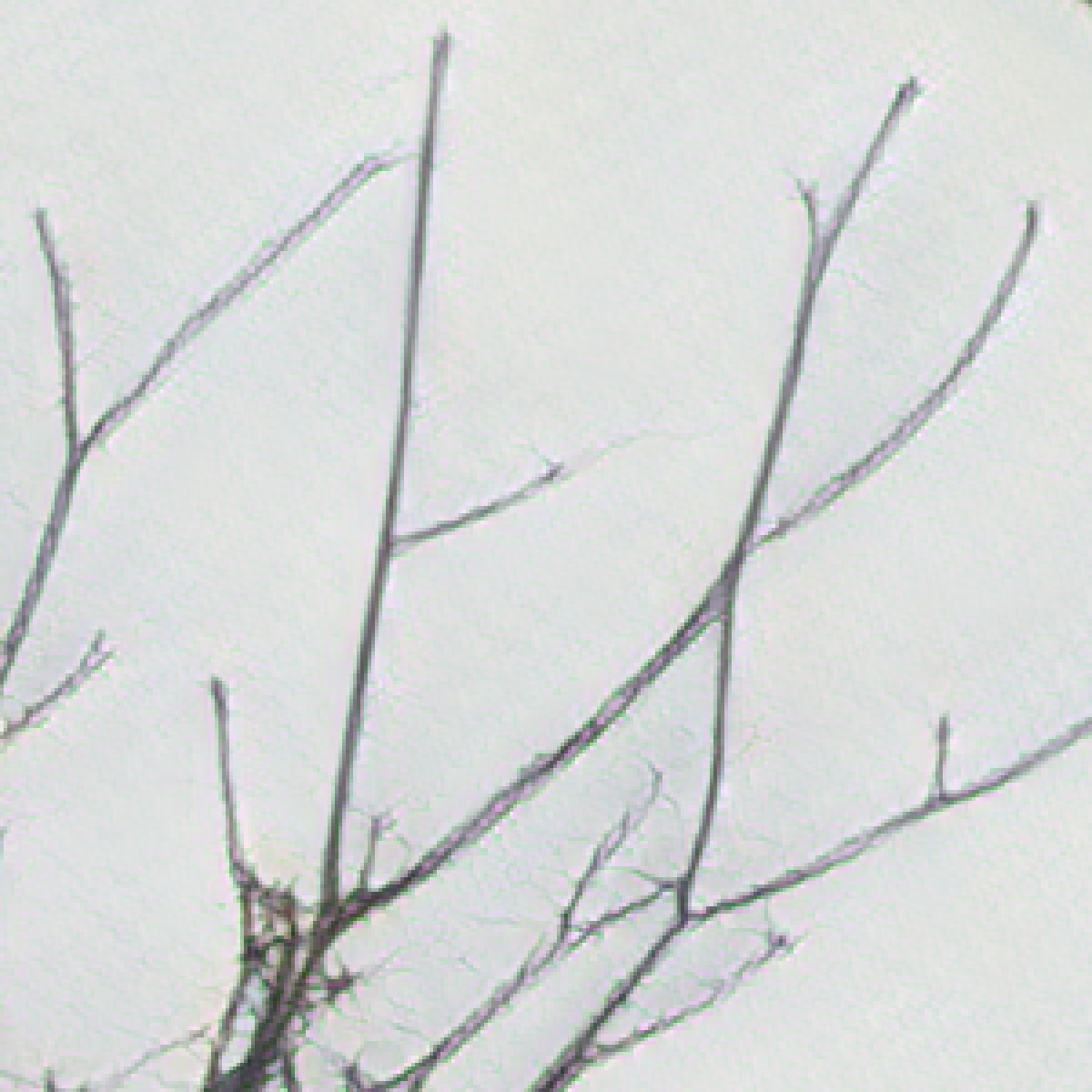}
\includegraphics[width=0.24\linewidth]{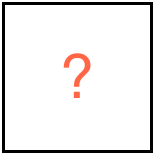}

\includegraphics[width=0.24\linewidth]{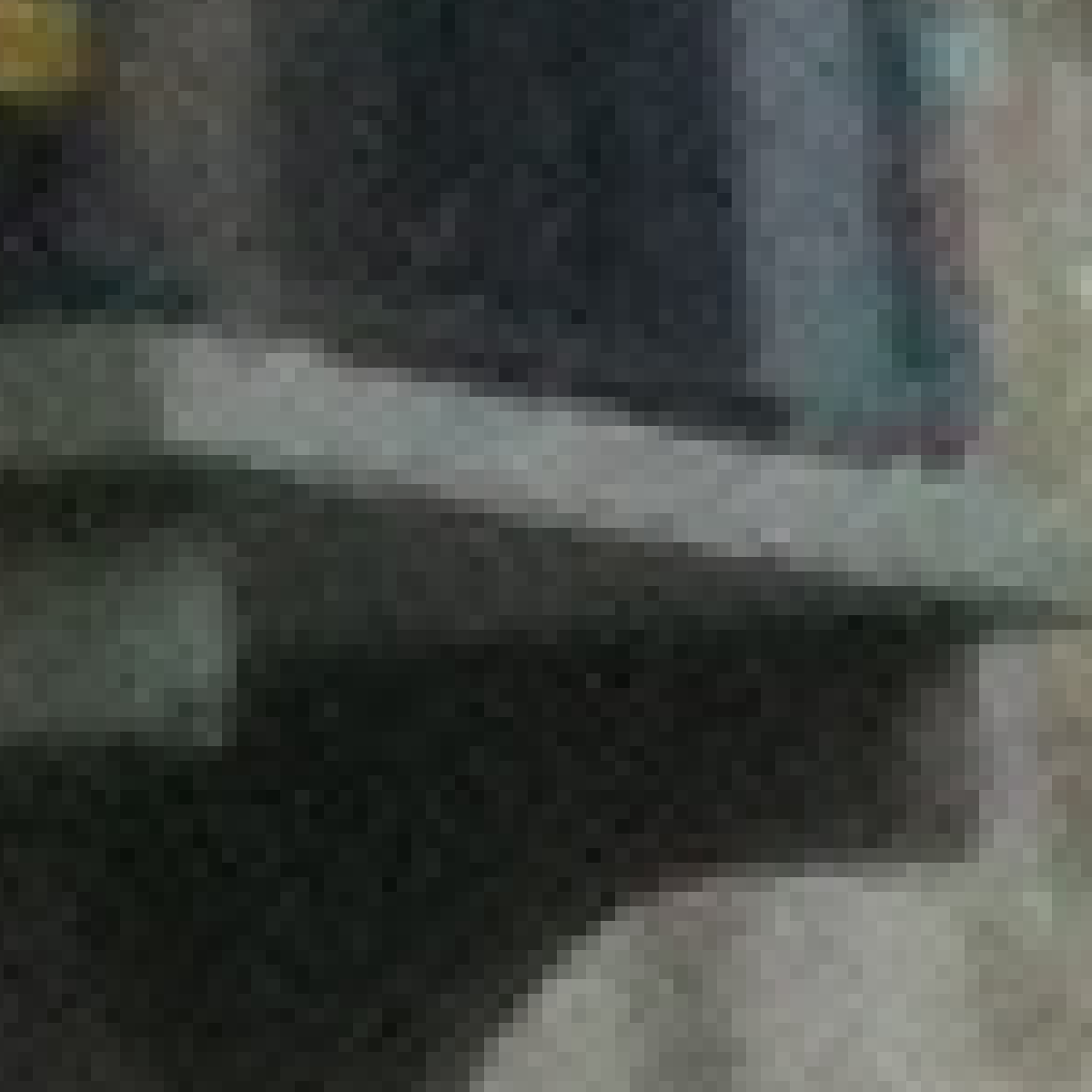}
\includegraphics[width=0.24\linewidth]{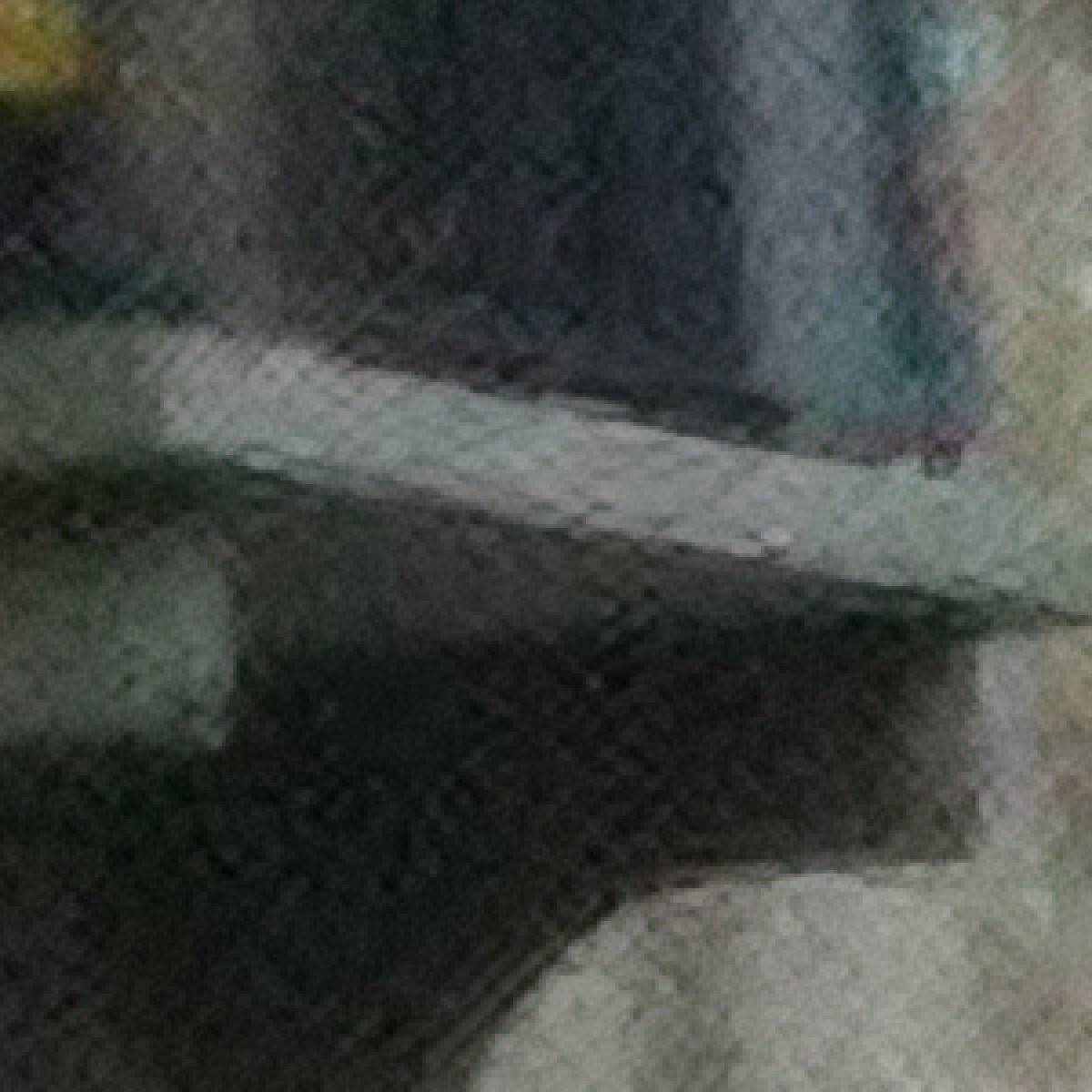}
\includegraphics[width=0.24\linewidth]{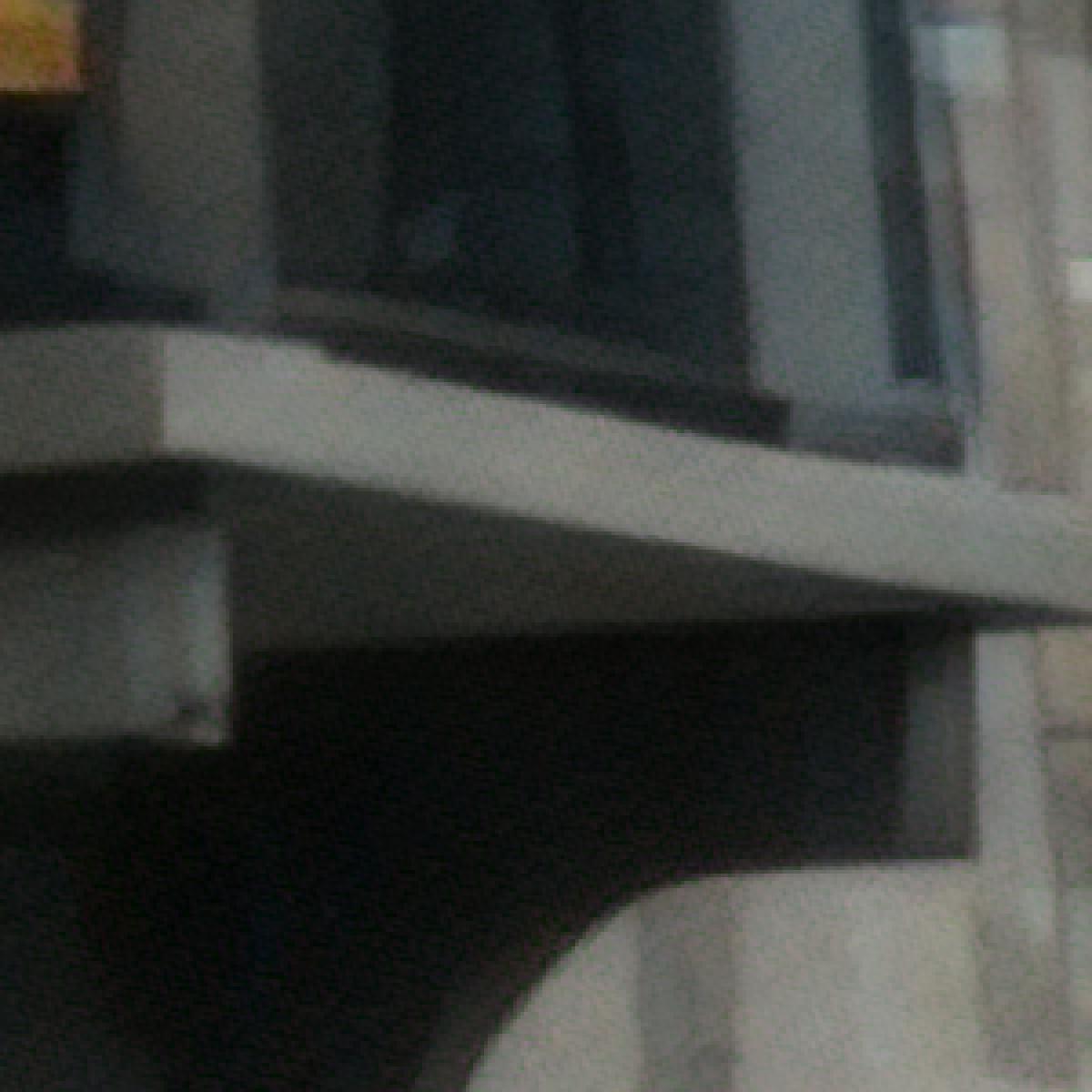}
\includegraphics[width=0.24\linewidth]{teaser/Questionmark.pdf}

\includegraphics[width=0.24\linewidth]{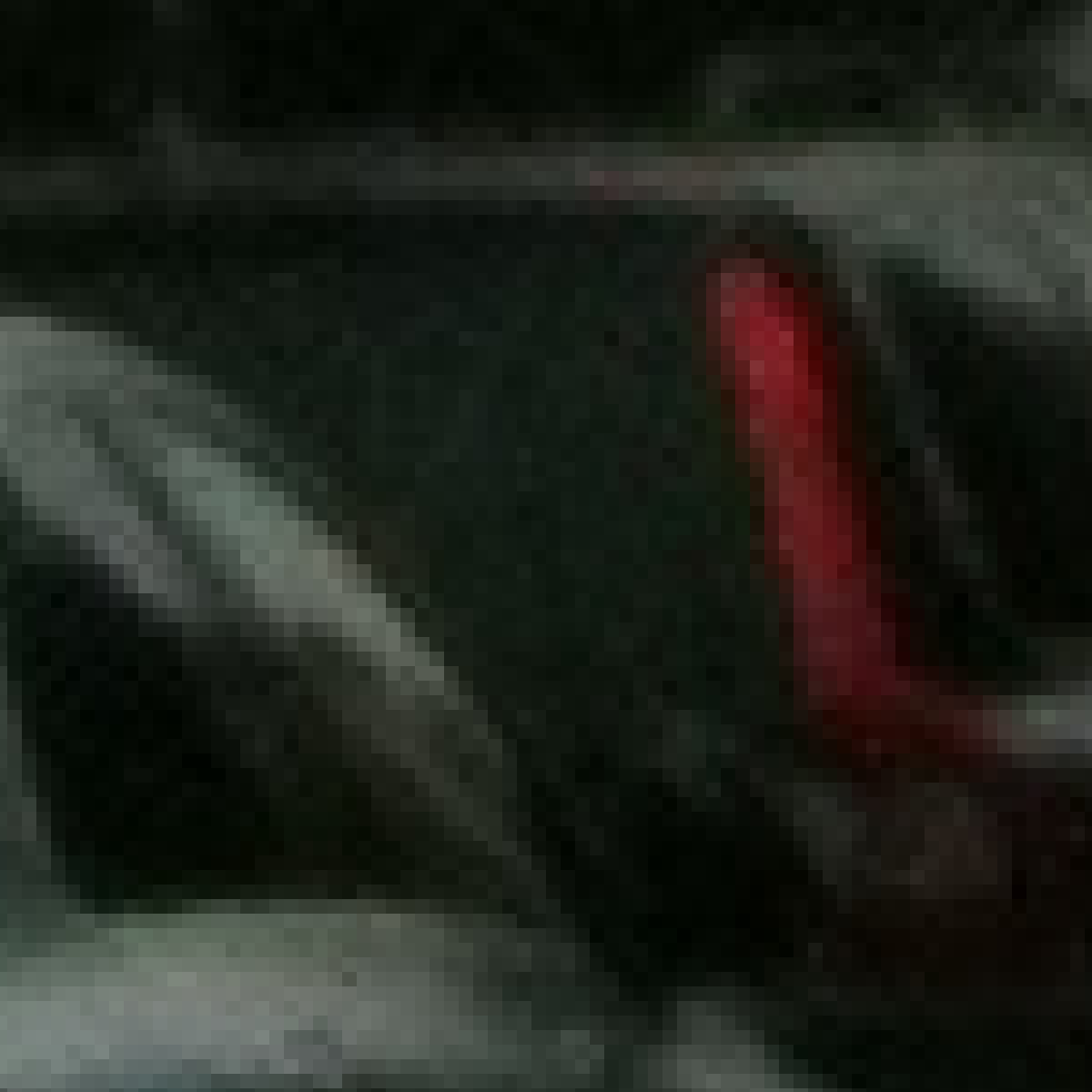}
\includegraphics[width=0.24\linewidth]{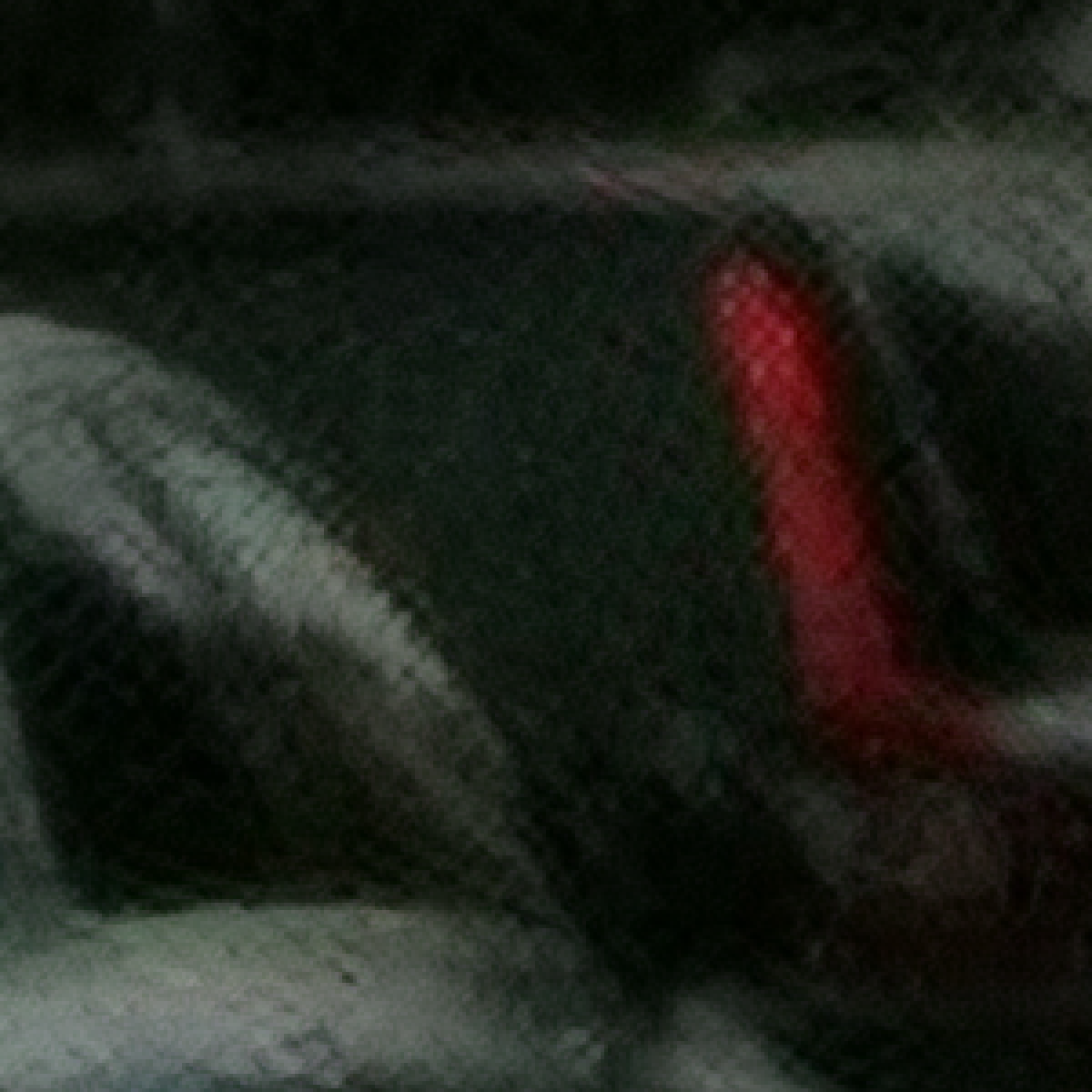}
\includegraphics[width=0.24\linewidth]{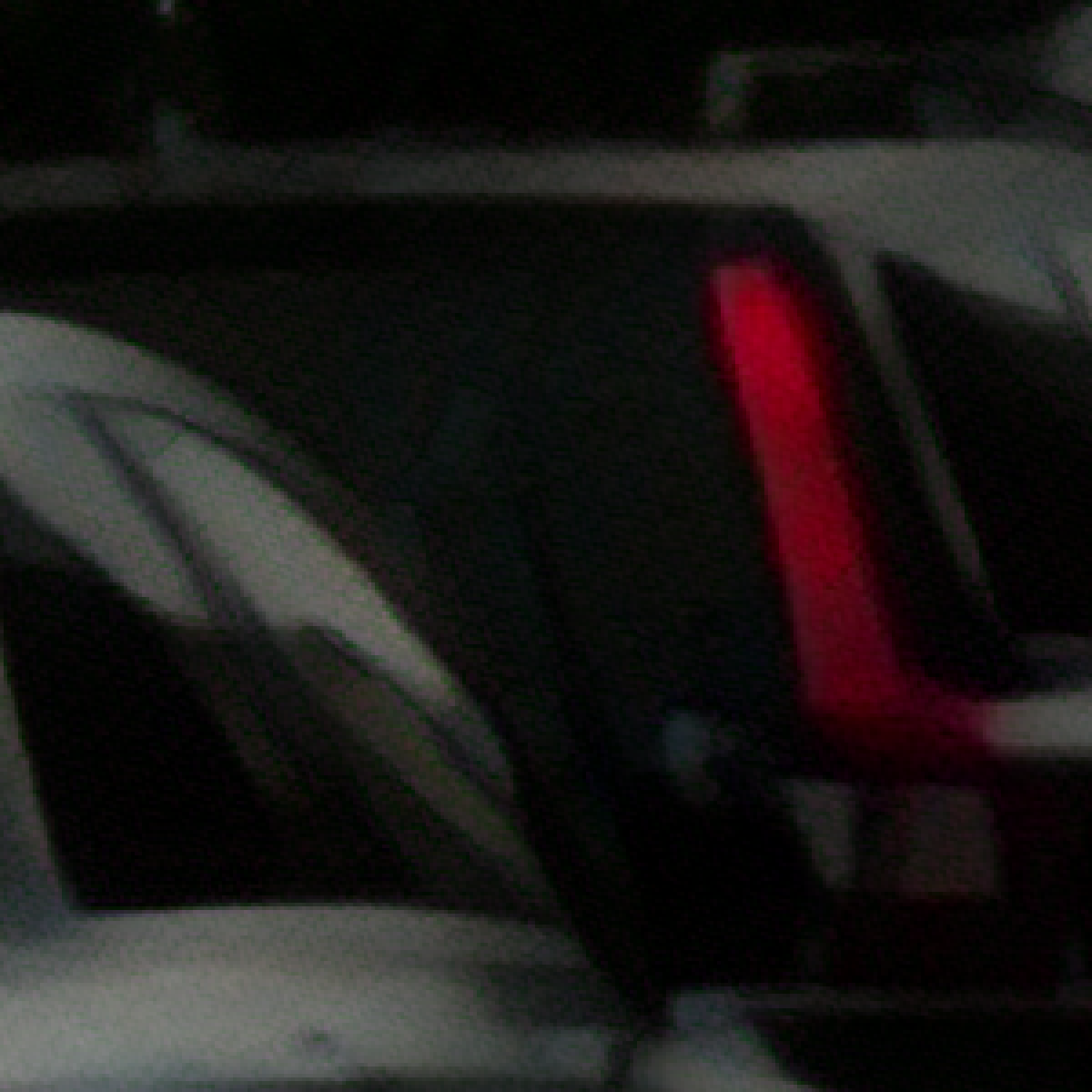}
\includegraphics[width=0.24\linewidth]{teaser/Questionmark.pdf}

\includegraphics[width=0.24\linewidth]{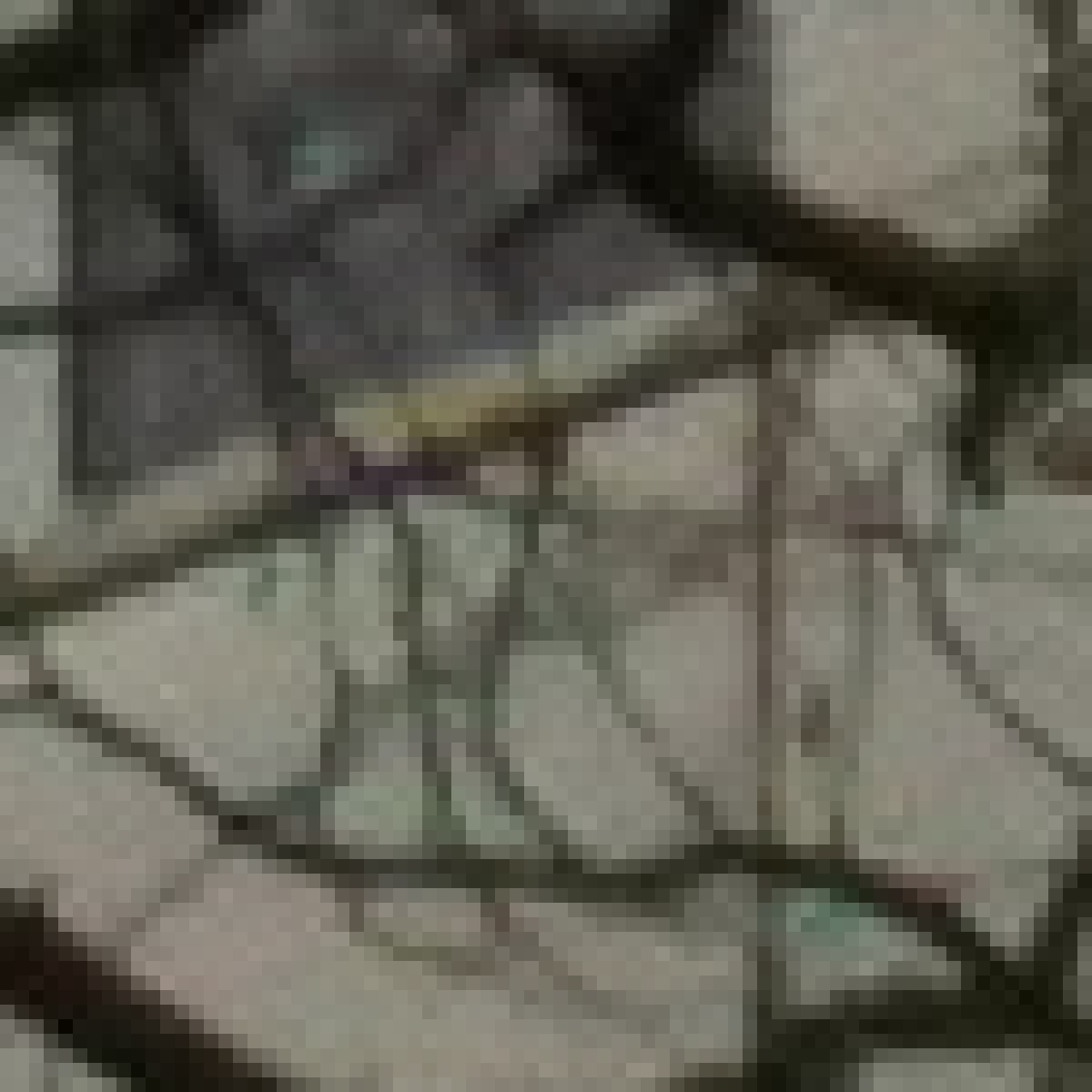}
\includegraphics[width=0.24\linewidth]{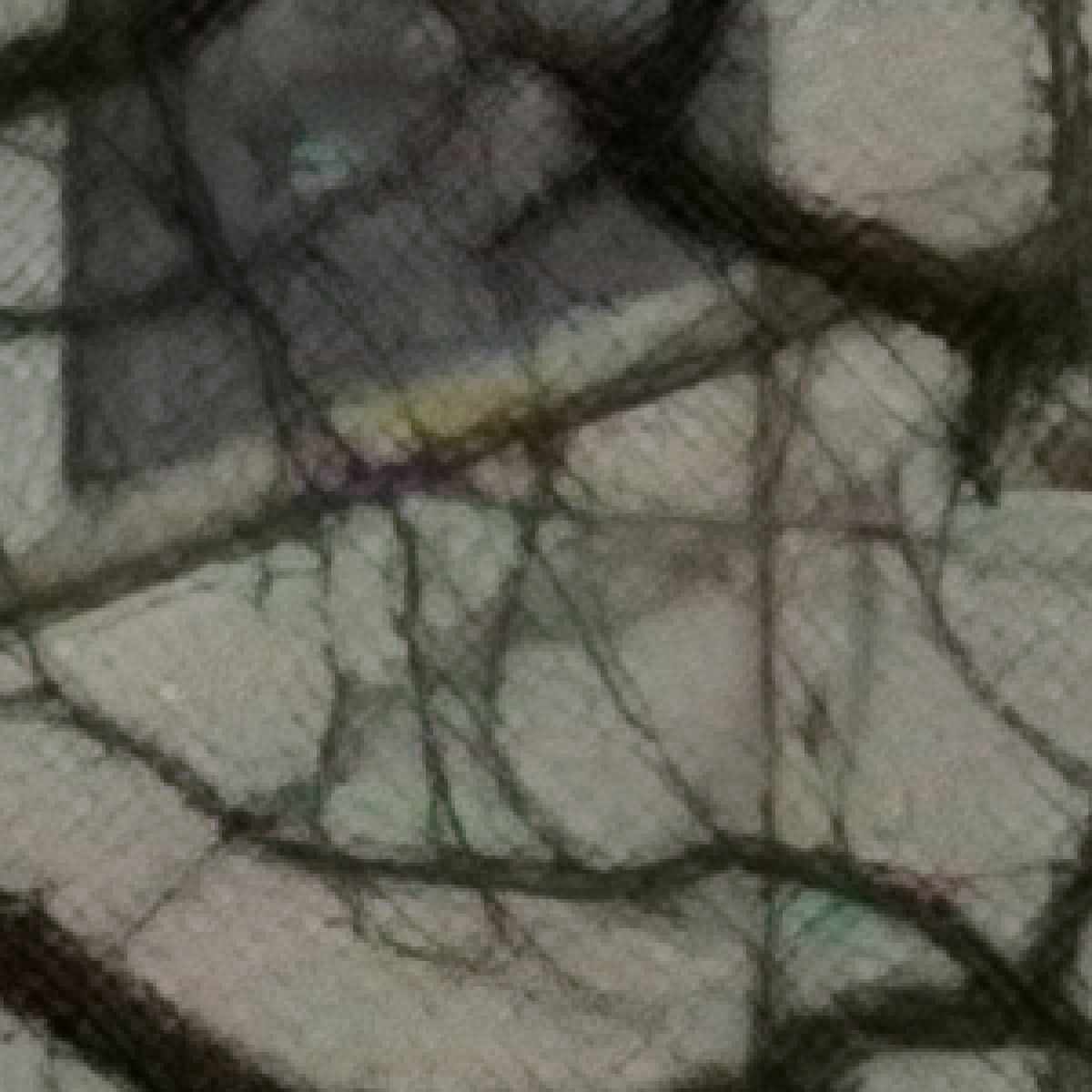}
\includegraphics[width=0.24\linewidth]{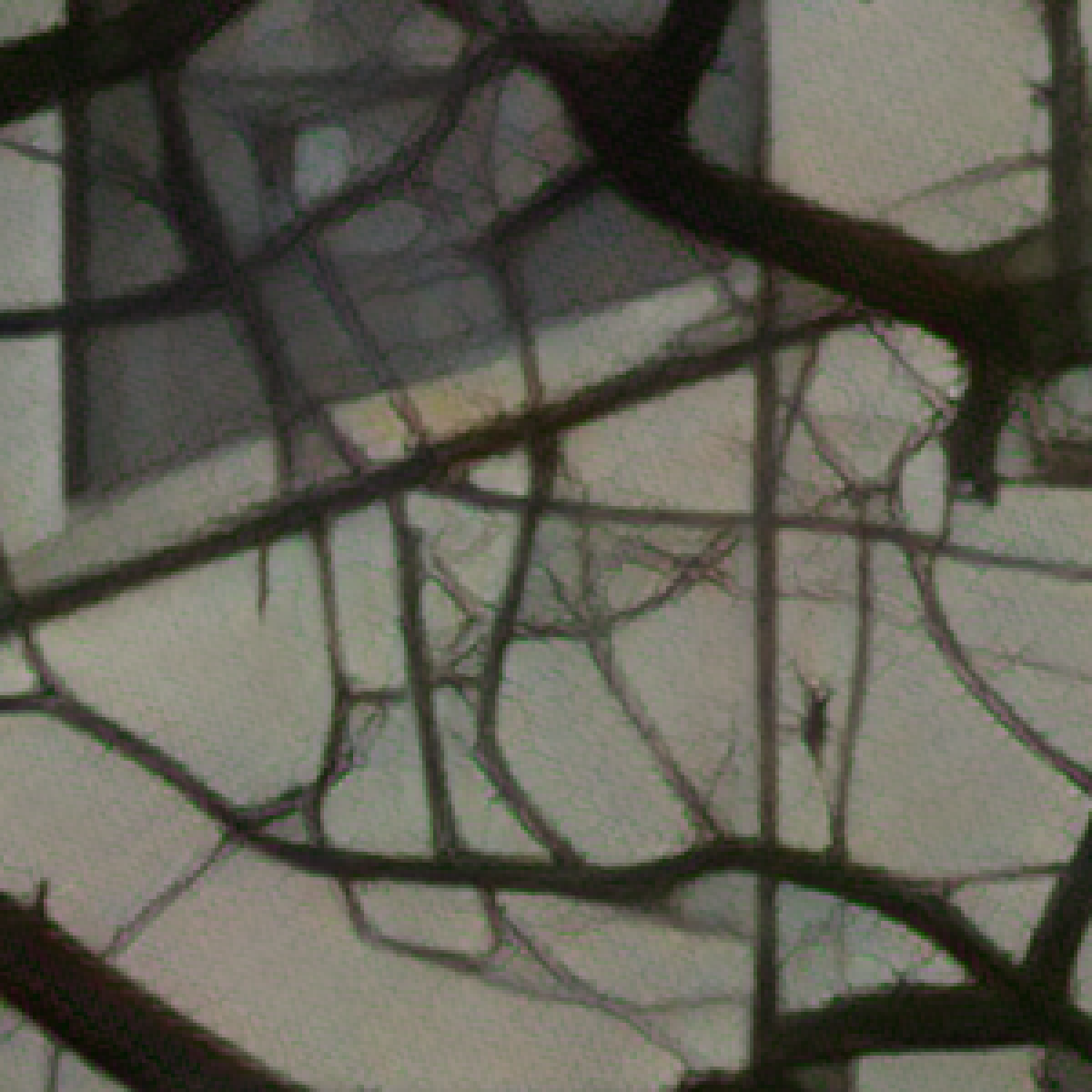}
\includegraphics[width=0.24\linewidth]{teaser/Questionmark.pdf}\vspace{-1mm}
\resizebox{\linewidth}{!}{
    \begin{tabular}{ C{2cm} C{2cm} C{2cm} C{2cm} C{2cm} }
        Original & ESRGAN & Ours & Ground Truth
\end{tabular}
}
\vspace{-4mm}
\caption{Super-resolving ($\times4$) natural images (left) with ESRGAN~\cite{wang2018esrgan}, trained on bicubic LR images, and our unsupervised approach. Ground truth data is unavailable in the real-world setting. Our approach learns to handle sensor noise and other artifacts in natural images, while ESRGAN fails to generalize.}
\vspace{-5mm}
\label{fig:teaser}
\end{figure}

State-of-the-art methods trained only to reconstruct images artificially downsampled with a bicubic kernel, do not generalize to natural images. As visualized in Figure~\ref{fig:teaser}, even small levels of noise causes a network trained only on bicubic images, in this case ESRGAN \cite{wang2018esrgan}, to output significant artifacts. In fact, this is expected as deep learning methods are known to be sensitive to significant differences between the train and test distributions. The ESRGAN has not seen noisy input images during train-time due to the smoothing effects introduced by bicubic downsampling.

In this work, we present a novel way of training a generic method in order to overcome the challenges of real-world SR. We address the shift between training and testing distributions arising from the bicubic downsampling by learning the corresponding \emph{inverse} mapping operation. To this end, we train a mapping from the bicubic images to the distribution of real-world LR images. By employing cycle consistency losses~\cite{zhu2017unpaired}, we learn this mapping in a fully unsupervised manner. The learned network is applied on bicubically downsampled images to generate paired LR and HR images that follow the real-world distribution. This allows us to learn the SR network on a realistic dataset, unaffected by the bicubic shift. Furthermore, the SR network is trained with direct pixel-wise supervision in the HR domain, without the need of any paired ground-truth data. Visual results of our approach on natural images is shown in Figure~\ref{fig:teaser}.

Due to the unavailability of paired data, we introduce a protocol for benchmarking real-world SR methods, based on simulating natural degradations. We analyze our approach in two scenarios, namely Domain (DSR) and Clean Super-Resolution (CSR). In the former case, the real-world data distribution is defined by one set of natural images. However, our approach generalizes to the case when the real-world input and output distributions of the SR network are different. We therefore introduce the CSR task, where the goal is to achieve a \emph{clean} super-resolved image, defined by a separate output distribution of high-quality images. We demonstrate the effectiveness of our approach on the aforementioned benchmark, and compare it to baseline methods and state-of-the-art approaches. Finally, we show qualitative results for the task of super-resolving real-world smartphone images on the DPED~\cite{ignatov2017dslr} dataset.

\section{Related Work}

Until very recently, single image super-resolution (SISR) methods were primarily benchmarked in terms of PSNR, for the task of super-resolving bicubic downsampled images. While traditionally addressed with classical techniques \cite{irani1991improving, freeman2002example, park2003super, yang2010image, huang2015single}, current approaches \cite{dong2014learning, dong2016image, kim2016accurate, lai2017deep, lim2017EDSR, fan2017balanced, ahn2018fast, ahn2018image, haris2018deep, huang2018densely} employ deep learning methodologies to train a mapping from LR to HR.
Among the latter, EDSR~\cite{lim2017EDSR} notably introduced a ResNet inspired architecture, better adapted for the task at hand. For training the network however, these methods rely on the L1 or L2 losses. While these losses are closely related to the PSNR evaluation metric, they do not preserve the natural image characteristics, generally leading to a blurry result~\cite{zhang2018unreasonable}. To address this problem, Ledig et al.~\cite{ledig2017photo} introduced an objective function aimed at
perceptually more pleasing results. The novel objectives were a GAN discriminator and a loss computed in the VGG feature space. While providing inferior PSNR compared to state-of-the-art, the super-resolved images experienced significantly better perceptual quality. Following this philosophy, the recent winner of the PIRM2018~\cite{ignatov2018pirm} challenge ESRGAN~\cite{wang2018esrgan}, proposed further architectural improvements to further enhance the perceptual quality.

Despite their success, the aforementioned approaches are severely limited by their reliance on the bicubic downsampling operation for training data generation. This operation eliminates most high frequency components and therefore, significantly altering the natural image characteristics, such as noise, compression artifacts, and other corruptions. The bicubic assumption therefore rarely reflects the real-world scenario. \emph{Blind} SR generalizes the problem by assuming LR and HR image pairs with an unknown degradation and downsampling kernel. 
Early attempts~\cite{begin2004blind} to this problem include explicitly estimating the unknown point spread function itself~\cite{michaeli2013nonparametric, gu2019blind}.
Another direction of research aims to completely remove the need for external training data by performing \emph{image-specific} SR. Following this idea, ZSSR~\cite{shocher2018zero} trains a lightweight network using only the testing image itself, by performing extensive data augmentation. However, this approach still employs a fix downsampling operation to generate synthetic pairs at test time. Furthermore, the image-specific learning leads to extremely slow prediction.

A few recent works address the \emph{unsupervised} SR setting, where no paired LR-HR pairs are given and the relation between LR and HR images is unknown.  The Cycle-in-Cycle network~\cite{yuan2018unsupervised} learns a mapping from the original input image to a \emph{clean} image space, using a framework that employs cycle consistency losses.  The SR network itself is trained by only employing indirect supervision in the LR domain, in addition to the usual perceptual GAN-discriminator. In contrast, our framework allows direct supervision in the HR domain, resulting in better training of the SR network itself. Furthermore, instead of ``cleaning'' the input image during train and test time, we learn a mapping to the original input domain for only the training. Another work focuses on the downsampling process~\cite{kim2018task} in order to improve the SR. However, SR is only performed on
images with the learned downsampling operation, and is therefore not applicable to our real-world scenario. Also Bulat~\etal~\cite{bulat2018learn} focus on the problem of learning the downsampling process. However, this approach specifically addresses the problem of super-resolving faces, where strong content priors can be learned by the network. In contrast, we tackle the general SR problem, not putting any assumptions on the image content. Lastly, recent works \cite{zhang2019zoom, chen2019camera, Cai_2019_CVPR_Workshops} propose strategies to capture real LR-HR image pairs. However, these methods rely on complicated data collection procedures, requiring specialized hardware, that is difficult and expensive to scale. Our approach operates without the need of any additional data, greatly increasing its use and applicability.

\section{Proposed Method}

\subsection{The Super-Resolution Problem}

In essence, super-resolution (SR) is the problem of increasing the resolution of natural images. However, this problem comes with a fundamental challenge that has been largely ignored up until very recently. Namely, the lack of \emph{natural} LR and HR image pairs, which are needed for evaluation and training. Therefore, research in SR has long relied on the use of known downscaling operators (e.g. bicubic) in order to artificially generate a corresponding LR image pair. While this simplification has historically also served the development of SR methods, it is fundamentally limiting.

\begin{figure}[t]
\includegraphics[width=\linewidth]{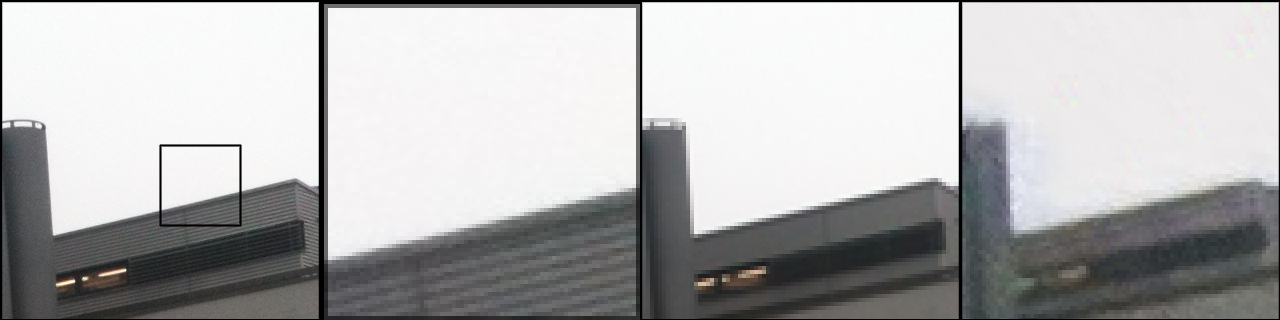}%
\vspace{-1mm}
\resizebox{\linewidth}{!}{\begin{tabular}{ C{2cm} C{2cm} C{2cm} C{2cm} }
    Original & Crop & Bicubic & Restored
\end{tabular}}
\vspace{-3.5mm}
\caption{Here we visualize the effects for bicubic downsampling and compare it to our domain distribution learning. The original HR image contains significant sensor noise, which almost completely disappears after bicubic downsampling. This is clearly observed when compared with the same-resolution crop of the original image. Our learned mapping $G$ restores the image characteristics present in the original image (right). }
\vspace{-4mm}
\label{fig:problem_with_bicubic}
\end{figure}

Bicubic downsampling can drastically change the natural characteristics of an image by, \eg, removing sensor noise and compression artefacts. A real-world example is shown in Figure~\ref{fig:problem_with_bicubic}. The natural image (left) is affected by natrual sensor noise. However, the corresponding bicubically downsampled image does not preserve these characteristics. Hence, a network trained to super-resolve the latter image cannot be expected to generalize to the original real-world distribution.

To formalize the problem, we let $X$ denote the natural image we wish to super-resolve. We also introduce the distribution $p_X$ of such natural images $X \sim p_X$ on which we want our SR approach to operate. In practice, $p_X$ could be defined as images obtained from a specific camera or a dataset of real-world images. The aim is to learn a function $S$ that maps an image $X \sim p_X$ to a high resolution image $\hat{Y} = S(X)$ that is distributed according to the output distribution $p_Y$. In applications, we could have $p_Y = p_X$, meaning that we want the characteristics of the image to remain unchanged after super-resolution. We term this setting domain-specific super-resolution (DSR). Another alternative would be to let $p_Y$ be defined by a set of high-quality images, which we call clean super-resolution (CSR) setting.

For most real-world applications it is incredibly hard and strenuous to collect natural image pairs $(X, Y)$ for SR.  In classical SR this is addressed by artificially constructing the input image $Z = B(Y)$, where $B$ is the bicubic downsampling operation. The task is then aimed to super-resolve $Z$ to match the original image $S(Z) \approx Y$. However, as illustrated in Figure~\ref{fig:problem_with_bicubic}, the
bicubically downsampled images $Z \sim p_Z$ do not match the input distribution, \ie \ $p_Z \neq p_X$. Unfortunately, methods trained in this manner struggle when supplied with real data $X \sim p_X$.

Related to our discussion is the concept of \emph{blind} SR. In this setting, the input images $X$ are assumed to be generated from the output images $Y$ with some fixed and simple transformation that is unknown. Often, a more general downsampling kernel $k$ is used in combination with a non-linear degradation function $f_\text{deg}$, such that $X = f_\text{deg}(k * Y)$. Some methods try to find the kernel $k$ from data or learn the transformation end-to-end.

\begin{figure*}[t]
\begin{center}
\includegraphics[width=1\linewidth,trim={0 0 3.3cm 0},clip]{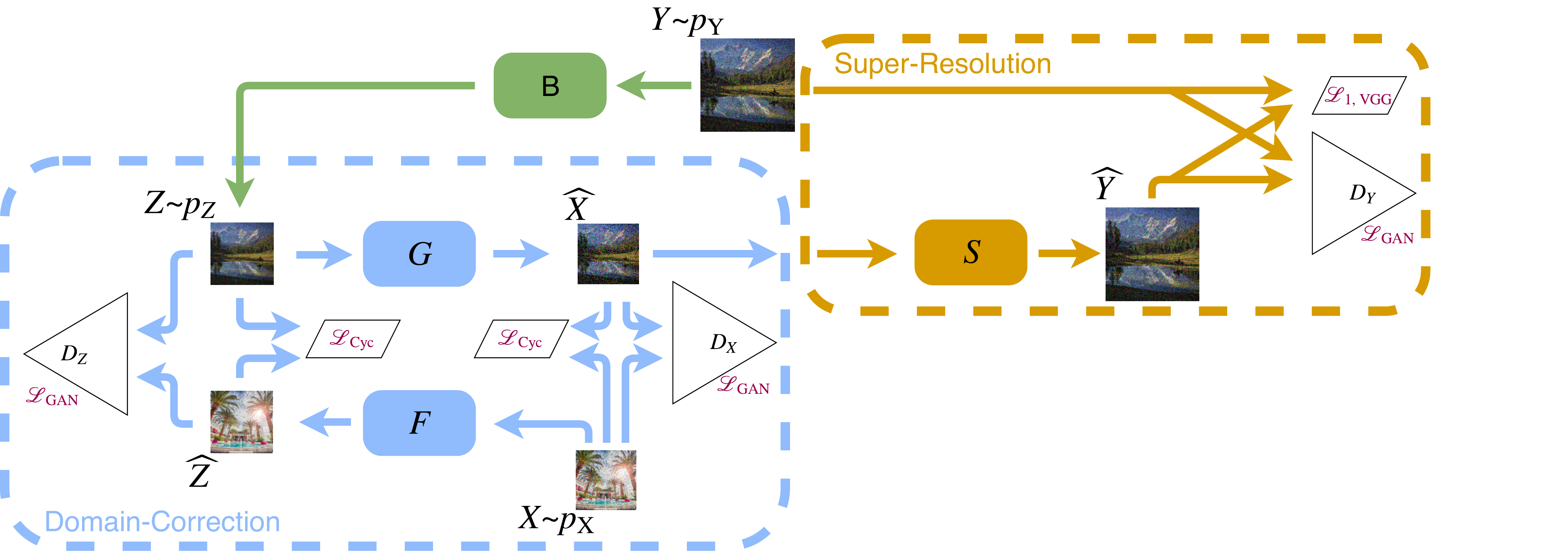}
\end{center}
\vspace*{-4mm}
    \caption{ Schematic overview of our approach. In the first step, we learn the domain distribution network $G$, depicted in blue. Given unpaired data from the input $p_X$ and output $p_Y$ distributions, the generator $G$ is trained in a GAN framework by employing cycle consistency losses. The SR network $S$ is trained in a second stage, depicted in orange, using pairs $(\hat{X}, Y)$ generated by our domain distribution network $G$.}
\vspace*{-4mm}
\label{fig:method-overview}
\end{figure*}

\subsection{Overview}

The real-world SR setting, addressed in this work, can be seen as a generalization of blind SR. In our approach, we assume no particular relation, such as a parameterized transformation, between the input $X$ and output $Y$ images. We only assume that a set of input image samples $\{X_i\}_{i=1}^M \sim p_X$ and a set of output image samples $\{Y_j\}_{j=1}^N \sim p_Y$ are available. These image samples are not paired. Given this data, the problem is to learn a mapping
$S$ that can super-resolve a new image $X \sim p_X$ such that $S(X) \sim p_Y$. In order to train $S$ from such unpaired data, we learn a function $\hat{X} = G(Z)$ that maps the bicubically downsampled image $Z = B(Y)$ from the output distribution to an image sample $\hat{X} \sim p_X$ that fits the input distribution $p_X$. This effectively constructs an input-output training pair $(\hat{X}, Y)$, allowing the SR network $S$ to be learned in a supervised manner such that $S(\hat{X}) \approx Y$. The main advantage of our approach is that the SR network can be trained with direct pixel-wise supervision in the HR domain. The proposed framework is depicted in Figure~\ref{fig:method-overview}. 

We first train the generator $G$, called the \emph{domain distribution network}, in a conditional GAN setting. This is performed by employing a discriminator network aiming to differentiate the generated images $\hat{X} = G(B(Y)), Y \sim p_Y$ from true input images $X \sim p_X$. Since no paired output is available, we enforce a cycle consistency loss by employing a second generator $F$ mapping input images $X$ to $\hat{Z} = F(X) \sim p_Z$. Crucially, we train the domain distribution network $G$
\emph{independently} from the SR network $S$. While, this may seem counter intuitive at first, it is clearly motivated from the fact that the networks $G$ and $S$ have fundamentally conflicting objectives. The aim of $G$ is to map a bicubically downsampled image $Z = B(Y)$ from the output distribution to an image $\hat{X}$ following the input distribution $p_X$, such that a faithful training sample $(\hat{X},Y)$ is generated for the SR network. The network $S$ simply aims to super-resolve any image from $p_X$. If both networks were to be trained jointly using the cycle-consistency loss for $S(G(B(Y))) \approx Y$, the networks $S$ and $G$ would collaborate in order to minimize the aforementioned loss. This leads to severe overfitting and poor generalization. As illustrated in Figure~\ref{fig:method-overview}, we train the SR network is a second separate training stage, using the training pairs generated by the network $G$.

\subsection{Domain Distribution Learning}
\label{sec:domaincorrection}

The task of the domain distribution learning $\hat{X} = G(Z)$ is to map a bicubic downsampled image from the output distribution $Z = B(Y)$ to the input distribution $p_X$. Since we do not have access to paired samples, we need to venture into unsupervised learning territories. We firstly employ a GAN discriminator $D_X$, tasked to differentiate between the generated $G(Z)$ and images drawn from the input distribution $p_X$. For this, we employ the original GAN formulation,
\begin{align}
\mathcal{L}_{GAN}(G, D_X) = \, & \mathbb{E}_{X \sim p_X}[ \log D_X(X)] \,+ \\ \nonumber
& \mathbb{E}_{Y \sim p_Y} [ \log (1 - D_X(G(B(Y)))] \,.
\label{eq:Cycle_GAN_loss}
\end{align}

To preserve the image content, despite the lack of paired images, we employ cycle consistency losses \cite{chu2017cyclegan}. A second generator $F$ is tasked to map images from the input domain $p_X$ to the domain of bicubic downsampled images $p_Z$, where $Z = B(Y)$. We then add cycle consistency losses as,
\begin{align}
\mathcal{L}_{cyc}(X,Y) = \, & \mathbb{E}_{Y \sim p_Y}[||F(G(B(Y))) - B(Y)||_1] \,+ \\ \nonumber
& \mathbb{E}_{X \sim p_X}[||G(F(X)) - X||_1] \,.
\label{eq:Cycle_Consistency_loss}
\end{align}
They constrain the generators $G$ and $F$ to be each others approximate inverses. Hence, the image shall be preserved if mapped through $G$ and then $F$ back to the original domain. Analogous to \eqref{eq:Cycle_GAN_loss}, we add a discriminator $D_Z$ and similar loss on the bicubic side. The full objective is thus,
\begin{align}
\mathcal{L}_{\text{DDL}}(G, F, D_X, D_Y) = \,&\mathcal{L}_{GAN}(G, D_X) \,+ \\ \nonumber
&\mathcal{L}_{GAN}(F, D_Z) \,+ \lambda \mathcal{L}_{cyc}(G,F).
\label{eq:Cycle_Objective}
\end{align}
The full architecture is shown in Figure~\ref{fig:method-overview} (blue). 

\noindent\textbf{Network architectures}
For our experiments we designed the domain distribution mapping $G$ based on the CycleGAN architecture~\cite{chu2017cyclegan}. The generators $G$ and $F$ use a ResNet architecture with nine blocks. We replace the transposed convolution layers with bi-linear upsampling followed by a standard convolution.  We found this to be beneficial for learning stability, and it effectively removed checkerboard pattern artifacts. Furthermore, we found the $\tanh$ non-linearity on the output to be harmful for color consistency, and therefore use no non-linear activation at the output. The discriminators $D_X$ and $D_Y$ consist of a three-layer network architecture that operate on a patch level \cite{li2016precomputed, isola2017image}.

\noindent\textbf{Training details}
We adopt the training procedure proposed in CycleGAN, using 200 epochs and the Adam optimizer with $\beta_1 = 0.5$. The starting learning rate is set to $0.0002$.

\subsection{Super-Resolution Learning}

Here we describe the learning of the SR network $S$. In the absence of paired ground-truth data, we train the network with pairs $(\hat{X}_j, Y_j)$, where the input image $\hat{X}_j = G(B(Y_j))$ is generated by our domain distribution network $G$. We employ the pixel-wise content loss \cite{lim2017EDSR},
\begin{equation}
\mathcal{L}_{1}(S) = \mathbb{E}_{Y \sim p_Y} \|S(\hat{X}) - Y \|_1 \,.
\end{equation}
Following the success of SRGAN~\cite{ledig2017photo}, we also employ the VGG feature loss, that is known to better correlate with perceptual quality
\begin{equation}
\mathcal{L}_{\text{VGG}}(S) = \mathbb{E}_{Y \sim p_Y} \|\phi(S(\hat{X})) - \phi(Y)\|^2_2 \,.
\end{equation}
Here, $\phi$ denotes the feature activations extracted from the VGG network. We extract the features at the same depth as SRGAN, which is after the activation of the 4th convolutional layer, before the 5th maxpooling layer.

For better perceptual quality, we further employ a GAN discriminator $D_Y$. To this end, we adopt the relativistic discriminator employed in ESRGAN \cite{wang2018esrgan}. As opposed to the conventional discriminator, providing an absolute real/fake probability for each image, a relative score real/fake is estimated compared to a set of real of fake images.
\begin{equation}
    D_Y(Y, \hat{Y})(C) = \sigma(C(Y) - \mathbb{E}[C(\hat{Y})]))
\end{equation}
Where $C(Y)$ is the raw discriminator output and $\sigma$ is the sigmoid function. The SR network is trained with added perceptual loss,
\begin{align}
\mathcal{L}_{\text{RaGAN}}(S,C) = & -\mathbb{E}_{Y\sim p_Y}[\log(1 - D_{Y}(Y, S(X)))] \nonumber \\ 
& - \mathbb{E}_{X\sim p_X}[\log(D_{Y}(S(X), Y))]\,.
\label{eq:relativistic_G}
\end{align}
This results in the total loss of
\begin{equation}
\mathcal{L}_{\text{}}(S, D) = \mathcal{L}_{\text{VGG}}(S) + \lambda \mathcal{L}_{\text{RaGAN}}(S,C) + \eta \mathcal{L}_1(S) \,.
\label{eq:relativistic_G}
\end{equation}
The GAN loss is multiplied by a weight $\lambda$, balancing the guidance of the two pixel-wise losses $\mathcal{L}_{\text{VGG}}$ and $\mathcal{L}_{\text{RaGAN}}$ against the GAN loss $\mathcal{L}_{\text{GAN}}$.

\noindent\textbf{Network architecture}
Our approach is agnostic to the specific architecture of the SR network $S$. For simplicity, we adopt the recently proposed ESRGAN architecture, which is the winner of the PIRM 2018 challenge~\cite{ignatov2018pirm}. It introduced a new building block called Residual-in-Residual Dense Blocks, improving stability of training. We augment the ESRGAN network with a final color adjustment layer, to ensure a faithful reproduction of the color palette in the input LR image. This layer adjusts the local mean RGB value to that of the low-resolution image.

\noindent\textbf{Training details}
To train our SR network, we start from pre-trained ESRGAN~\cite{wang2018esrgan} generator and discriminator networks. We then perform 50000 training iterations. We use that ADAM optimizer with an initial learning rate of $10^{-4}$ and set $\beta_1 = 0.9$ and $beta_2 = 0.999$ for both the Generator and Discriminator. We use the learning rate schedule in \cite{wang2018esrgan}, decreasing it by a factor of 0.5 after 10\%, 20\%, 40\% and 60\% of the total number of iterations. 

\section{Experiments}

In this section, we present comprehensive quantitative and qualitative evaluation of our approach. We first discuss the setup and datasets employed in our experiments. Detailed results are provided in the supplementary material.

\subsection{Experimental Setup}
\label{sse:experimental_setup}

\begin{figure}[t]
\vspace{-0.2cm}
\begin{center}
\includegraphics[trim={0 0 0cm 0},width=1\linewidth,clip]{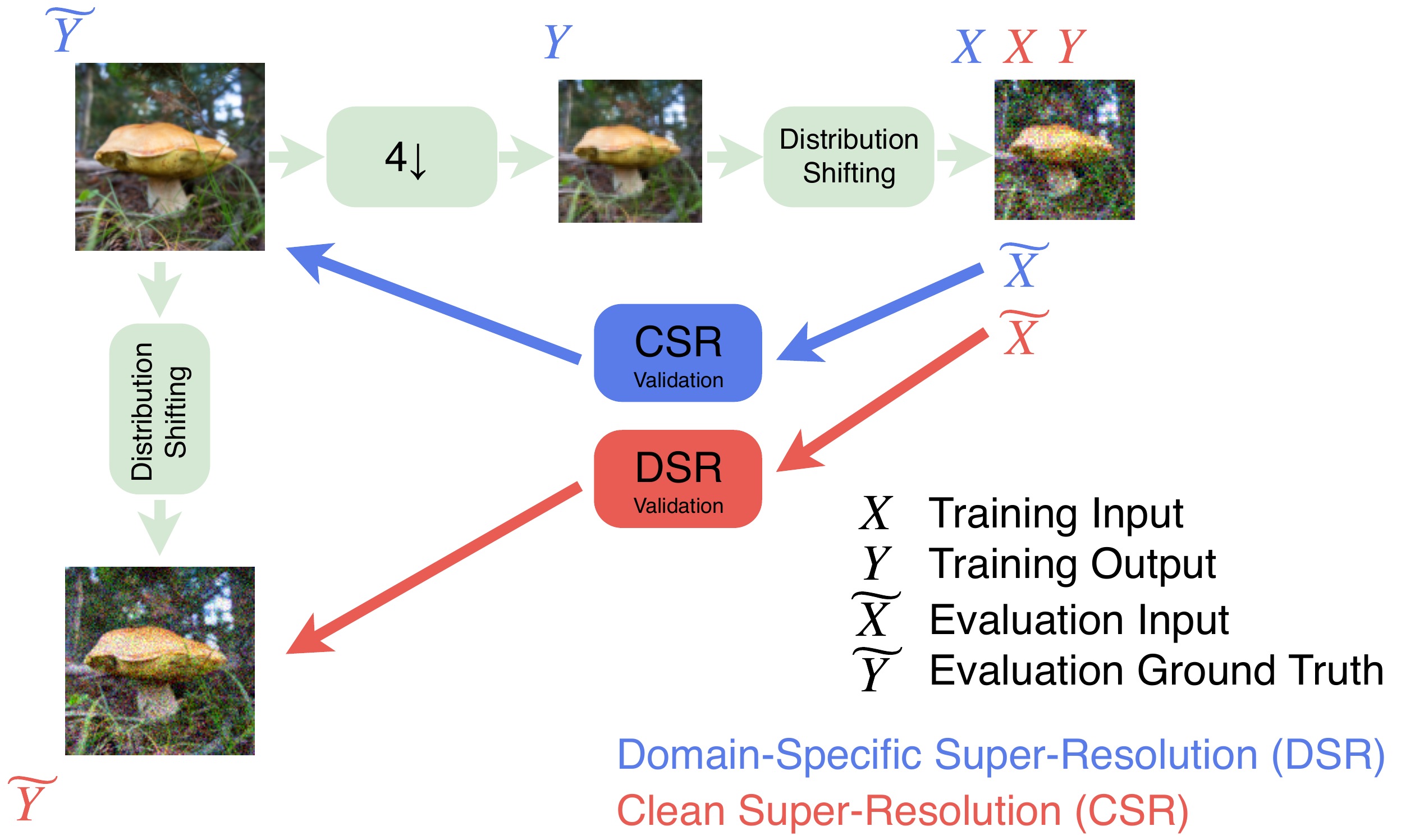}
\end{center}
\vspace*{-4mm}
\caption{Overview of our data generation procedure for benchmarking unsupervised SR methods. See text for details.}
\label{fig:data-pipeline}
\vspace*{-4mm}
\end{figure}

We present a novel strategy for evaluating real-world SR methods. In traditional SR the bicubic downsampled image $Z = B(Y)$ is super-resolved and compared to $Y$. In the real-world scenario, we do not have access to a ground-truth image, complicating quantitative analysis. On the other hand, we can closely simulate the real-world scenario by constructing the sets input $\{X_i\}_{i=1}^M \sim p_X$ and output $\{Y_j\}_{j=1}^N \sim p_Y$ training images by applying downscaling and synthetic degradations to a dataset of original images. The type of degradation is \emph{unknown} to the SR approach. For evaluation, we further generate a set of ground truth pairs $(\tilde{X}_i, \tilde{Y}_i)$, where $\tilde{X}_i \sim p_X$ and $\tilde{Y}_i \sim p_Y$. These are inaccessible to the network during training, and only used for evaluation purposes. We consider two scenarios: DSR and CSR, detailed below.

\textbf{Domain-Specific Super-Resolution (DSR)} The input $X$ and target $Y$ images shares the same real-world distribution, \ie \ $p_X = p_Y$. Thus, the aim is to produce super-resolved images $\hat{Y}$ that are of the same distribution as its input images. The training set is generated by first downsampling the image and then simulating the real-world degradation. In the DSR case the same training set $\{X_i\}_{i=1}^M$ represents both the input $p_X$ and output $p_Y$ distribution. For evaluation, the input image $\tilde{X}_i$ is constructed using the same procedure as for the training images $X_i$. The corresponding ground-truth output image $\tilde{Y}_i$ is obtained by directly adding the degradation to the original HR image. The procedure is visualized in Figure~\ref{fig:data-pipeline}.

In \textbf{Clean Super-Resolution (CSR)}, the goal is to super-resolve an input image $X$ such that the output image $\hat{Y}$ fits another distribution $p_Y$. We let $p_Y$ be defined by a dataset of high quality images. Therefore, we employ the unaltered original image from the dataset to be the ground-truth output $\tilde{Y}_i$. The corresponding LR image $\tilde{X}_i$ used for the evaluation is generated as in the DSR case above. We also employ the same training set of input
images $\{X_i\}_{i=1}^M$ as for the DSR. In the CSR case however, the output training data $\{Y_j\}_{j=1}^N$ represent a different distribution of clean images. These are generated by bicubically downsampling the original image. The resulting image thus represents a clean ideal output from the SR network. See Figure~\ref{fig:data-pipeline} for a schematic description of the procedure.

\textbf{Degradations}
To model the real-world setting we evaluate unsupervised SR approaches using two types of image degradations: JPEG compression artifacts and simulated sensor noise. In case of JPEG artifacts we use a quality setting of 30. JPEG compression artifacts are a common when applying super-resolution to images captured by smartphones or acquired from the internet. In the second case, we employ white Gaussian noise with a standard deviation of $\sigma = 8$. This simulates the case of real-world sensor noise present in \eg, \ low light conditions or small sensor sizes. 

\begin{figure}
\resizebox{\linewidth}{!}{\begin{tabular}{ C{3cm} C{3cm} C{3cm} }
    PSNR = \textbf{20.81} & PSNR = $\infty$ & PSNR = 18.90 \\
    LPIPS = 0.7889 & LPIPS = 0 & LPIPS = \textbf{0.2737}
\end{tabular}}
\includegraphics[width=0.32\linewidth]{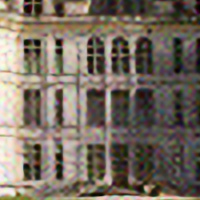}
\includegraphics[width=0.32\linewidth]{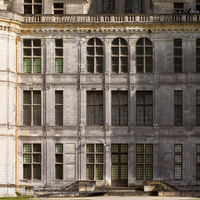}
\includegraphics[width=0.32\linewidth]{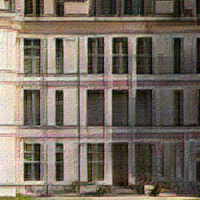}%
\vspace*{-1mm}
\resizebox{\linewidth}{!}{\begin{tabular}{ C{2cm} C{2cm} C{2cm} }
    EDSR & GT & Ours
\end{tabular}}
\vspace*{-4mm}
\caption{While the PSNR is only dependend on the pixel-wise distance between GT and the prediction, LPIPS is a more elaborate measure that takes the perceptual quality into account. Although the EDSR image is perceptually much worse than the prediction of our method, it scores higher in PSNR. The LPIPS distance is $56\%$ smaller for our method, which is perceptually superior.}
\vspace*{-4mm}
\label{fig:psnr_vs_lpips}
\end{figure}

\textbf{Quantitative Evaluation Measures} In order to quantitatively compare the different approaches we use the distance metrics PSNR, SSIM and LPIPS. While PSNR and SSIM are handcrafted methods, LPIPS is a learned metric for perceptual similarity~\cite{zhang2018unreasonable} between two images. In Figure~\ref{fig:psnr_vs_lpips} we provide a comparison of PSNR and the LPIPS, measures using the model provided by the authors.

\textbf{Datasets}
We use the DF2K~\cite{wang2018esrgan} dataset that was introduced for learning the ESRGAN. It is a merge of the DIV2K~\cite{timofte2017ntire} with 800 and the Flickr2K~\cite{timofte2017ntire} dataset with 2640 images. The mean size of DF2K is 1439x1935. We also perform experiments on the DPED~\cite{ignatov2017dslr} dataset, acquired by a smartphone camera. It contains natural images
with real-world sensor noise and other effects.

\begin{table}
	\newcommand{\sep}{~~}
	\resizebox{\linewidth}{!}
	{
		\begin{tabular}{@{}l@{~~~}l@{\sep}|@{\sep}c@{\sep}c@{\sep}c@{\sep}|@{\sep}c@{\sep}c@{\sep}c@{}}
                & & \multicolumn{3}{c}{Sensor Noise} & \multicolumn{3}{c}{JPEG Artifacts} \\
                      & Method & $\uparrow$PSNR & $\uparrow$SSIM & $\downarrow$ LPIPS &  $\uparrow$PSNR & $\uparrow$SSIM & $\downarrow$ LPIPS \\
                      \hline
                      \hline
                      \multirow{4}{1mm}{\rotatebox{90}{\resizebox{14mm}{!}{\textbf{DSR} Unsuperv.}}}
                      & Baseline              & 18.46 & 0.23 & 0.8182 & 22.22 & 0.41 & 0.6319 \\
                      & Cleaning the input    & 19.81 & 0.32 & 0.5737 & 22.23 & 0.56 & 0.4295 \\
                      & Low res.\ supervision & 21.46 & 0.36 & 0.4363 & 20.00 & 0.49 & 0.4483 \\
                      & \textbf{Ours}         & 22.43 & 0.40 & 0.2897 & 23.30 & 0.62 & 0.3732 \\
                      \hline
                      & DSR Supervised (ref.) & 23.97 & 0.48 & 0.1778 & 22.97 & 0.59 & 0.3526 \\
                      \hline
                      \hline
                      \multirow{4}{1mm}{\rotatebox{90}{\resizebox{14mm}{!}{\textbf{CSR} Unsuperv.}}}
                      & Baseline              & 18.48 & 0.23 & 0.7532 & 23.15 & 0.60 & 0.4887 \\
                      & Cleaning the input    & 22.10 & 0.55 & 0.4516 & 20.28 & 0.46 & 0.4889 \\
                      & Low res.\ supervision & 22.03 & 0.55 & 0.4401 & 20.16 & 0.49 & 0.4752 \\
                      & \textbf{Ours}         & 22.42 & 0.55 & 0.3645 & 22.80 & 0.57 & 0.3729 \\
                      \hline
                      & CSR Supervised (ref.) & 25.54 & 0.70 & 0.2103 & 22.60 & 0.58 & 0.3484 \\
		\end{tabular}
	}
	\vspace*{-1mm}
	\caption{Comparison of the four different versions in our ablation study for the DSR (upper half) and CSR (lower half) setting. We validate the methods on the DIV2K dataset. We also compare with training the SR network with full supervision, providing an upper bound for unsupervised methods. Our approach achieves superior perceptual quality measured by the LPIPS distance for sensor noise and JPEG artifacts.}
	\vspace*{-4mm}
	\label{tab:ablation}
\end{table}

\subsection{Ablation Study}
For our ablation study we use the DF2K dataset as training data and the validation image from DIV2K to measure the performance of the different methods. The quantitative comparison is done using the PSNR, SSIM and LPIPS measures. For comparisons, we mainly consider the LPIPS distance due to its higher correlation with perceptual similarity.

\begin{figure}
	
	\rotatebox[origin=c]{90}{\resizebox{4.5cm}{!}{
			\begin{tabular}{ C{3cm} C{3cm} }
				DSR - JPEG & DSR - Sensor Noise
			\end{tabular}
	}}%
	\begin{minipage}{\linewidth}%
	\includegraphics[width=0.955\linewidth]{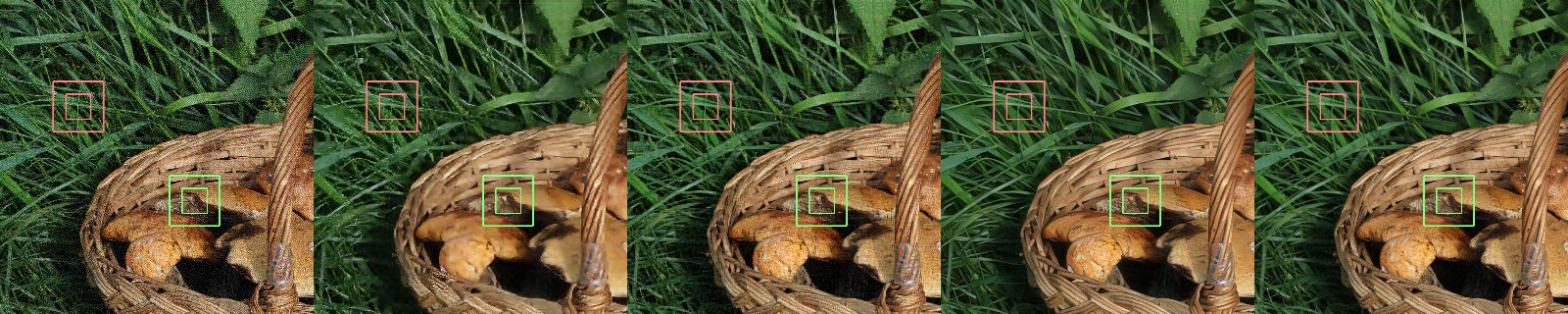}\vspace{-0.25mm}
	\includegraphics[width=0.955\linewidth]{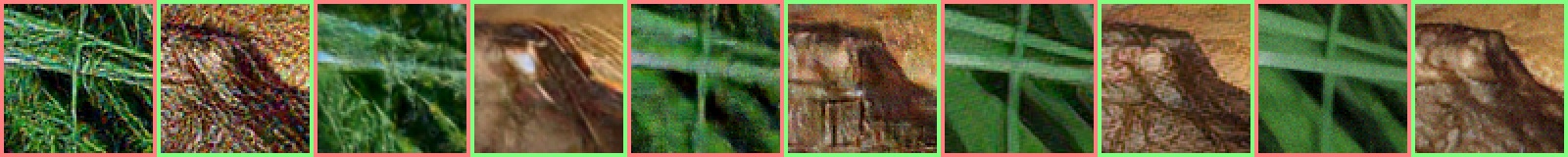}\vspace{-0.25mm}
	\includegraphics[width=0.955\linewidth]{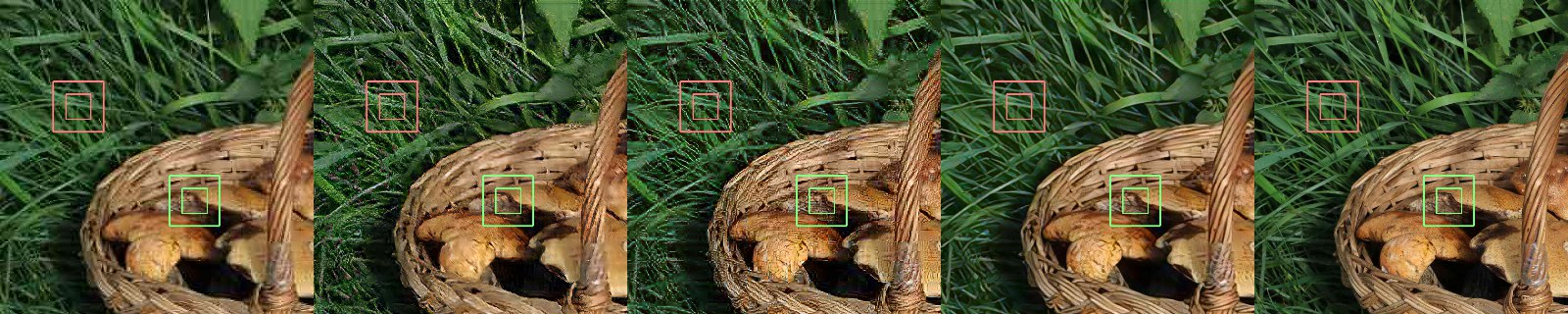}\vspace{-0.25mm}
	\includegraphics[width=0.955\linewidth]{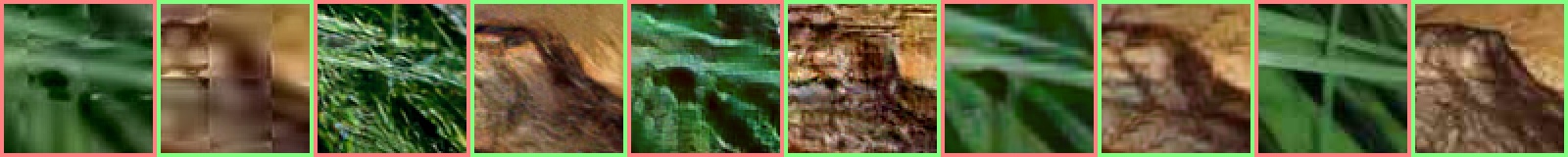}\vspace{-0.25mm}
	\end{minipage}
	
	\rotatebox[origin=c]{90}{\resizebox{4.5cm}{!}{
			\begin{tabular}{ C{3cm} C{3cm} }
				CSR - JPEG & CSR - Sensor Noise
			\end{tabular}
	}}%
	\begin{minipage}{\linewidth}%
	\includegraphics[width=0.955\linewidth]{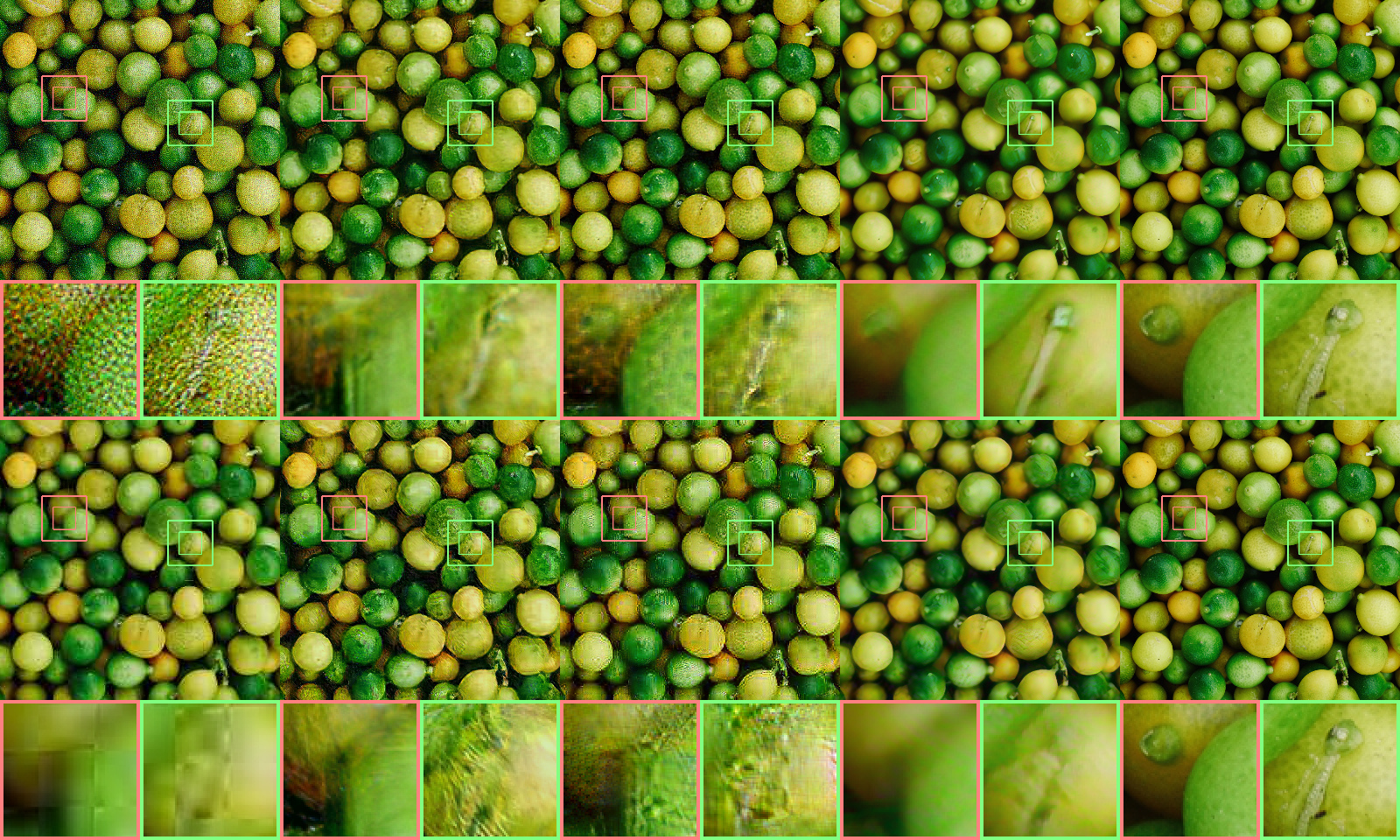}
	\end{minipage}
	
	\resizebox{\linewidth}{!}{
		\begin{tabular}{ C{0.2cm} C{2cm} C{2cm} C{2.5cm} C{2cm} C{2cm} }
			& Baseline & Cleaning the input & Low resolution supervision & \textbf{Ours} & GT
		\end{tabular}
	}
	\vspace*{-3.5mm}
	\caption{Ablation Study for DSR (Top) and CSR (Bottom) settings for four different versions using the DIV2K validation set. Our approach, in the second most right column, provides favorable results despite the strong noise or JPEG artifacts in the input images. The compared baseline variations generate strong artifacts in most cases. In the CSR JPEG case we observe a slightly blurred result compared to ground truth. This is due to the loss of detail caused by the JPEG compression.
	}
	\vspace*{-4mm}
	\label{fig:ablation}
\end{figure}

We evaluate four different approaches for the DSR and CSR setting. All methods are trained using the same settings. For SR network we employ a pretrained ESRGAN model that is then fine-tuned for each method, as described below. Quantitative and visual results are shown in Table~\ref{tab:ablation} and Figure~\ref{fig:ablation} respectively.

\textbf{Baseline} First, we compare with the standard approach of training the network on LR images generated by bicubic downsampling. For this purpose, we finetune the ESRGAN using image pairs $(Z, Y)$. In the case of sensor noise, the baseline achieves significantly inferior performance for both DSR and CSR (Table~\ref{tab:ablation}) compared to ours. This is due to the smoothing behaviour of the bicubic downsampling. The baseline ESRGAN does thus not see appropriate levels of noise during training. This leads to severe artifacts in both the DSR and CSR case (Figure~\ref{fig:ablation}). Our approach also improves in the case of JPEG artifacts, leading to a $24\%$ improvement in LPIPS in the CSR case.

\textbf{Cleaning the input} Another strategy for tackling the shift between train and test distribution, caused by the bicubic downsampling, is to map the input image $X$ to the bicubic distribution $\hat{Z} = F(X) \sim p_Z$ before applying the SR network, as proposed in \cite{yuan2018unsupervised}. With this strategy, the SR network $S$ is trained using bicubic data, exactly as in the baseline setting discussed previously. During inference, the input image $X$ is super-resolved as $\hat{Y} = S(F(X))$. In fact, in our approach, we already train such a mapping $F$ to ensure cycle consistency in the training of domain correction network $G$ (Section~\ref{sec:domaincorrection}). Since our training is fully symmetric between the two domains, including the architectures of $F$ and $G$, we use the generator $F$ trained in our framework for a fair comparison.

In case of sensor noise, this version improve the results compared to the baseline method, suggesting that some of the domain shift problem is alleviated. However, our approach further improves the LPIPS distance by 50\% in the DSR and 19\% in the CSR case. This is partly due to the fact that the SR network acts directly on the input image $X$, while the mapping $F$ can introduce artifacts or remove information. Our approach also achieves a significant improvement in the JPEG case.

\textbf{Low resolution supervision} Here, we compare our approach with performing supervision in the LR domain. Similar to \cite{yuan2018unsupervised}, we add another generator network $H$ that maps the super-resolved image $\hat{Y}$ back to the original domain. We then perform the direct pixel-wise supervision in the LR domain \emph{instead}, using the same losses $\mathcal{L}_1$ and $\mathcal{L}_\text{VGG}$ to ensure $H(\hat{Y}) \approx X$. We observed that this approach  leads to stronger GAN hallucinations, as shown in Figure~\ref{fig:ablation}. This tendency is also observed in the quantitative results, obtaining significantly worse LPIPS and PSNR in all cases. This demonstrates the importance of direct HR supervision provided by our method.

\textbf{Fully supervised} To assess the performance of our approach, we compare with fully supervised training using paired samples, otherwise unavailable to the network. We generate paired data $(\tilde{X}_i, \tilde{Y}_i)$ using the same strategy employed for evaluation, \ie \ by applying the ground-truth degradation. The ESRGAN is then finetuned on this data directly. Note that the ground-truth degradation operation is unknown for all other methods in this comparison. We observe that our approach achieves performance much closer to this upper bound for both DSR and CSR. In particular in the case of JPEG artifacts, where our unsupervised method is only slightly worse than full supervision.

\subsection{State-of-the-art Comparison on DIV2K}

\begin{table}
        \newcommand{\sep}{~~}
        \resizebox{\linewidth}{!}
        {
                \begin{tabular}{@{}ll@{\sep}|@{\sep}c@{\sep}c@{\sep}c@{\sep}|@{\sep}c@{\sep}c@{\sep}c@{}}
                        && \multicolumn{3}{c}{Sensor Noise} & \multicolumn{3}{c}{JPEG Artifacts} \\
                        & Method        & $\uparrow$PSNR & $\uparrow$SSIM & $\downarrow$LPIPS & $\uparrow$PSNR & $\uparrow$SSIM & $\downarrow$LPIPS \\
                        \hline
                        \multirow{6}{1mm}{\rotatebox{90}{\resizebox{7mm}{!}{DSR}}}
                        & ZSSR          & 23.65 & 0.47 & 0.6925 & 24.20 & 0.65 & 0.4584 \\
                        & EDSR          & 23.39 & 0.44 & 0.7137 & 24.12 & 0.65 & 0.4525 \\
                        & ESRGAN        & 17.17 & 0.19 & 0.7363 & 22.90 & 0.61 & 0.4447 \\
                        & ESRGAN FT     & 18.08 & 0.22 & 0.6233 & 23.24 & 0.62 & 0.4129 \\
                        & \textbf{Ours} & 22.43 & 0.40 & 0.2897 & 23.30 & 0.62 & 0.3732 \\
                        \hline
                        \multirow{6}{1mm}{\rotatebox{90}{\resizebox{7mm}{!}{CSR}}}
                        & ZSSR          & 24.87 & 0.60 & 0.6466 & 23.92 & 0.64 & 0.5447 \\
                        & EDSR          & 24.46 & 0.53 & 0.6824 & 23.83 & 0.63 & 0.5454 \\
                        & ESRGAN        & 17.39 & 0.19 & 0.9434 & 22.67 & 0.59 & 0.5069 \\
                        & ESRGAN FT IN  & 18.35 & 0.23 & 0.7595 & 23.01 & 0.60 & 0.4903 \\
                        & ESRGAN FT OUT & 17.35 & 0.19 & 0.9040 & 22.82 & 0.59 & 0.5087 \\
                        & \textbf{Ours} & 22.42 & 0.55 & 0.3645 & 22.80 & 0.57 & 0.3729 \\
                \end{tabular}
        }
        \vspace*{0mm}
        \caption{Comparison to state-of-the-art super-resolution methods on the DIV2K dataset using the DSR and CSR setting. Our approach achieves the best perceptual results in both sensor noise and JPEG artifact case.}
        \label{tab:stateOfTheArt}
        \vspace*{-4mm}
\end{table}

In the following we compare our approach with other state-of-the-art methods: ZSSR~\cite{shocher2018zero}, EDSR~\cite{lim2017EDSR}, ESRGAN~\cite{wang2018esrgan}. Therefore we use the original code and trained models. The method ZSSR applies a Zero-Shot learning strategy, where weights are learned for each image individually. The EDSR trains a ResNet-based model without perceptual loss. ESRGAN applies the same SR architecture and perceptual losses as in our approach. Both EDSR and ESRGAN are trained using bicubic supervision.

We also report the results of finetuning the ESRGAN on the same training data employed by our approach. In the case of CSR, where two distinct training sets are available, we further compare with finetuning the ESRGAN on each of those. The method "ESRGAN FT IN" is trained on the set of input images, while "ESRGAN FT OUT" is trained on the set of output images. In all cases we follow the same training procedure as~\cite{wang2018esrgan}, constructing the corresponding LR image using bicubic downsampling.

We evaluate the aforementioned approaches on the DIV2K validation set using the DSR and CSR setting as described in Section~\ref{sse:experimental_setup}. Results for the DSR and CSR settings are reported in Table~\ref{tab:stateOfTheArt}. 

\begin{figure}[t]
	\newcommand{\wid}{.095\linewidth}
	\begin{minipage}{0.03\linewidth}%
		\resizebox{2mm}{!}{\rotatebox{90}{
				\begin{tabular}{ C{3cm} C{3cm} }
					JPEG & Sensor Noise
				\end{tabular}%
		}}%
	\end{minipage}%
	\begin{minipage}{0.19\linewidth}%
		\includegraphics[width=\linewidth]{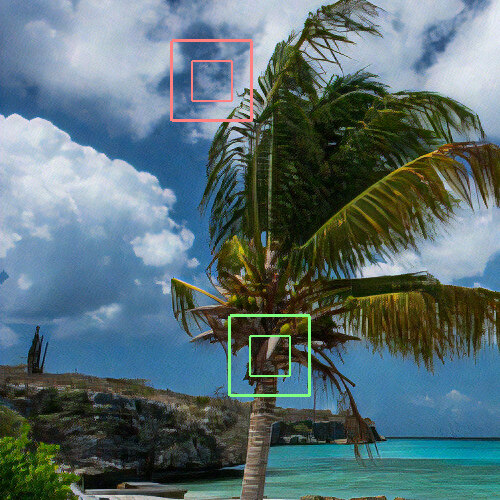}
		\includegraphics[width=\linewidth]{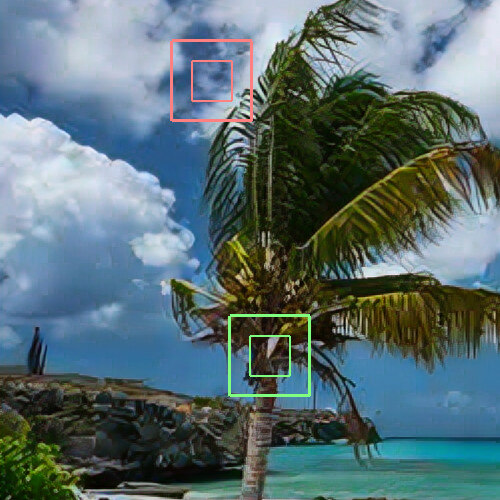}
	\end{minipage}%
	\begin{minipage}{\wid}%
		\includegraphics[width=\linewidth]{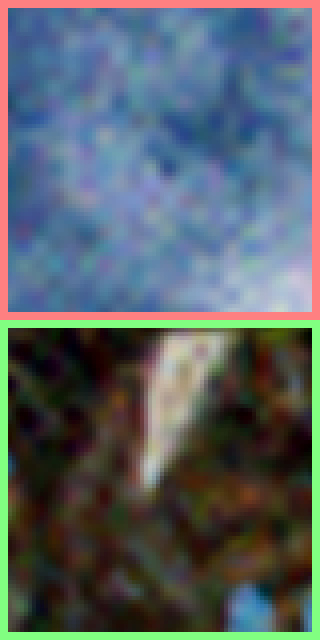}
		\includegraphics[width=\linewidth]{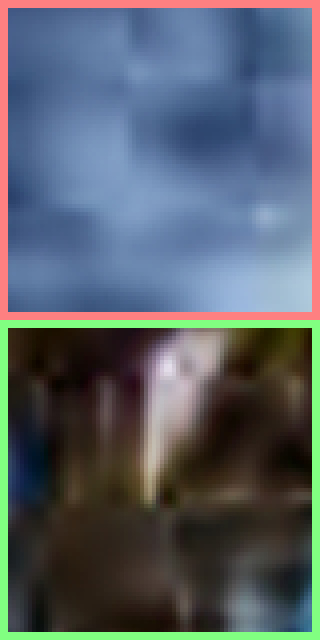}
	\end{minipage}%
	\begin{minipage}{\wid}%
		\includegraphics[width=\linewidth]{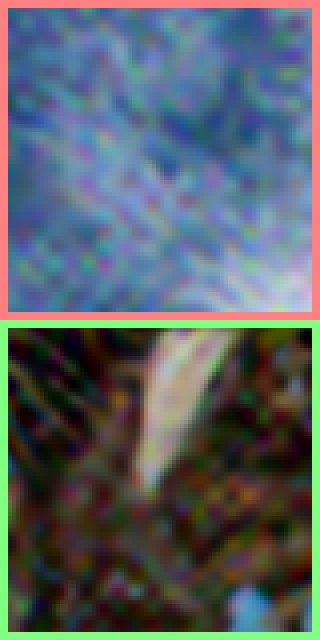}
		\includegraphics[width=\linewidth]{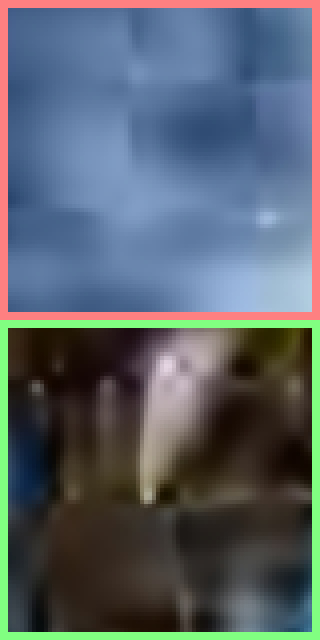}
	\end{minipage}%
	\begin{minipage}{\wid}%
		\includegraphics[width=\linewidth]{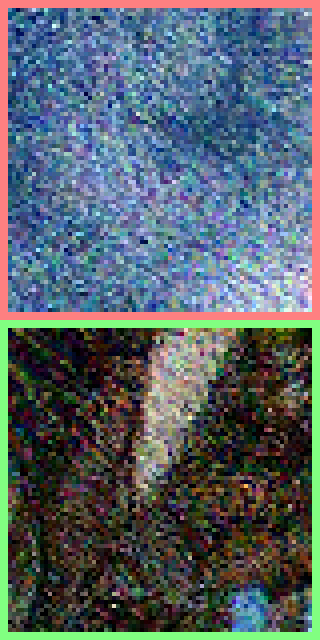}
		\includegraphics[width=\linewidth]{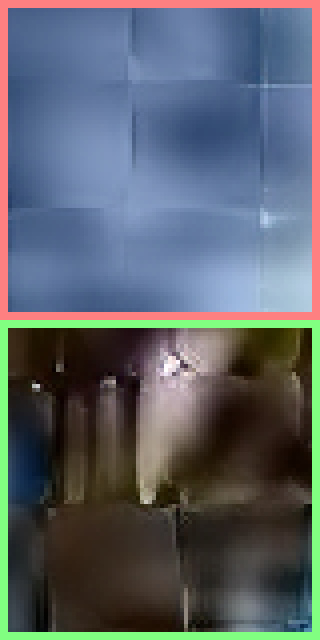}
	\end{minipage}%
	\begin{minipage}{\wid}%
		\includegraphics[width=\linewidth]{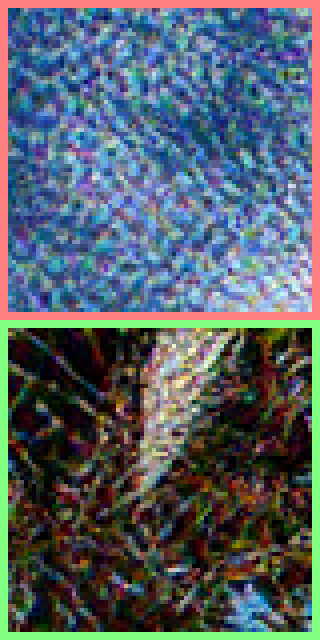}
		\includegraphics[width=\linewidth]{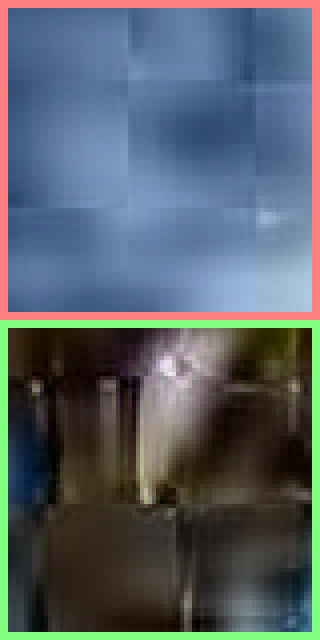}
	\end{minipage}%
	\begin{minipage}{\wid}%
		\includegraphics[width=\linewidth]{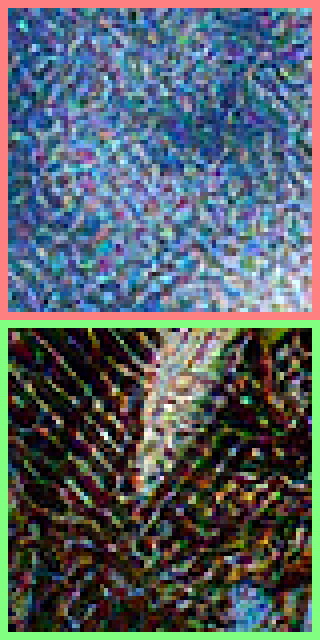}
		\includegraphics[width=\linewidth]{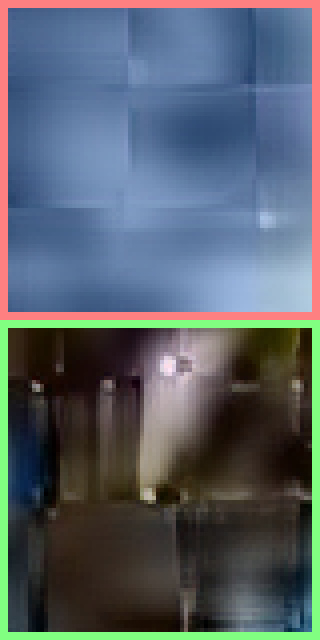}
	\end{minipage}%
	\begin{minipage}{\wid}%
		\includegraphics[width=\linewidth]{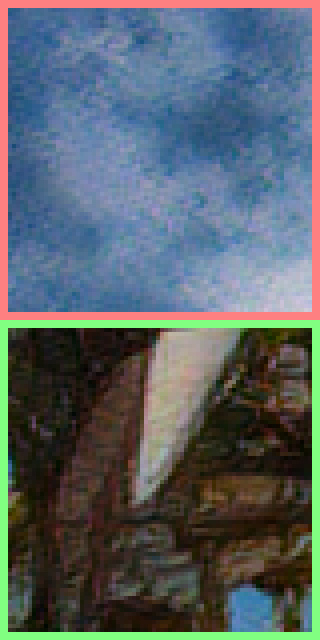}
		\includegraphics[width=\linewidth]{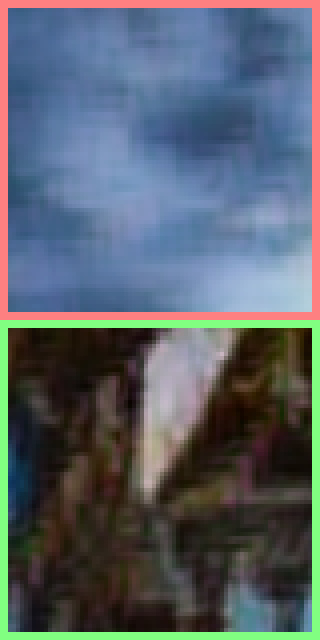}
	\end{minipage}%
	\begin{minipage}{\wid}%
		\includegraphics[width=\linewidth]{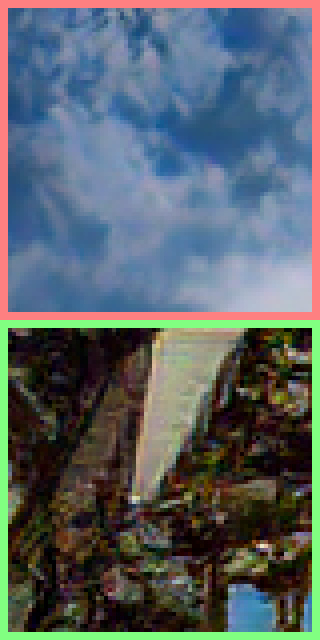}
		\includegraphics[width=\linewidth]{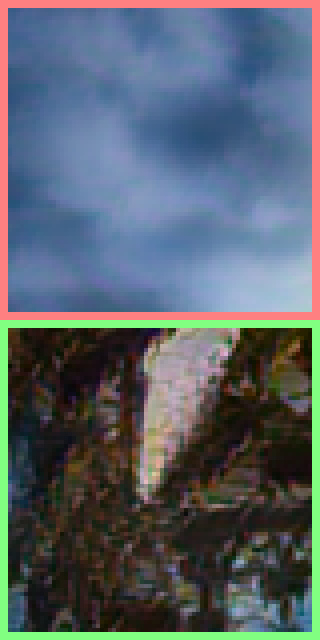}
	\end{minipage}%
	\begin{minipage}{\wid}%
		\includegraphics[width=\linewidth]{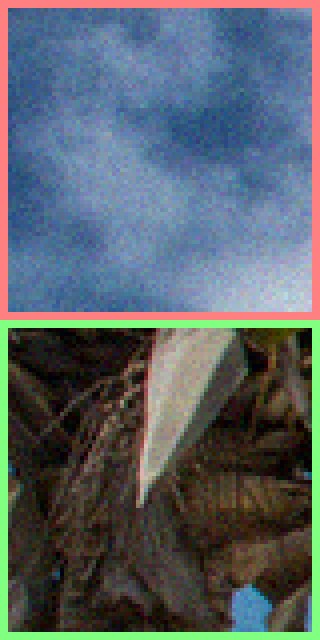}
		\includegraphics[width=\linewidth]{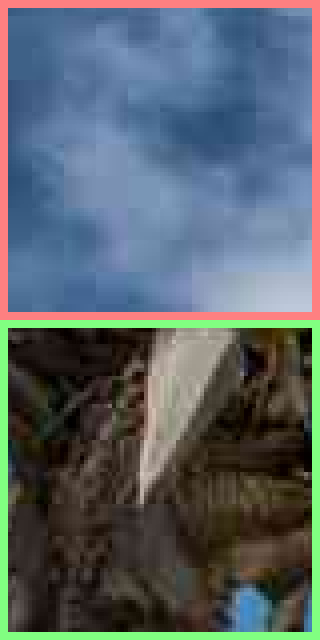}
	\end{minipage}%
	
	\newcommand{\sep}{~}
	\newcommand{\sz}{1.9cm}
	\resizebox{\linewidth}{!}{
		\begin{tabular}{@{}C{0.7cm}@{\sep}C{3.6cm}@{\sep}C{\sz}@{\sep}C{\sz}@{\sep}C{\sz}@{\sep}C{\sz}@{\sep}C{\sz}@{\sep}C{\sz}@{\sep}C{\sz}@{\sep}C{\sz} }
			& \textbf{Ours} (DSR) & ZSSR & EDSR & ESRGAN & ESRGAN FT IN & ESRGAN FT OUT & \textbf{Ours} (DSR) & \textbf{Ours} (CSR) & GT
		\end{tabular}
	}
	\vspace{0mm}
	\caption{Qualitative comparison between our approach and four state-of-the-art approaches. Our approach is able to recover the original distribution. This is due to the input and output domain awareness that the other methods lack of.}
	\vspace{-4mm}
	\label{fig:stateOfTheArt_div2k}
\end{figure}

\noindent\textbf{Sensor Noise}
In the case of Sensor noise, the ESRGAN provides poor perceptual quality with LPIPS distances of $0.7363$ and $0.9434$ for DSR and CSR. Finetuning the model on the given data only provides minor improvements. This is due to the domain distribution shift caused by the bicubic downsampling leading to clean input images during training. EDSR experiences better robustness to noisy input with LPIPS values of $0.7137$ and $0.6824$ for DSR and CSR respectively. Among previous approaches ZSSR achieves the best performance, owing to its zero-shot learning strategy. Our unsupervised approach achieves the best overall perceptual quality, significantly reducing the LPIPS error metric by $58\%$ and $44\%$ for DSR and CSR respectively, in the sensor noise setting.

As show in Figure~\ref{fig:stateOfTheArt_div2k} the ESRGAN approaches produce strong artifacts in the case of noisy input. This is likely due to the perceptual loss that encourages the network to output high-frequency components in order to provide a sharp image. In contrast, our method do not suffer from such artifacts despite employing a strong perceptual loss. This demonstrates that our SR network has learned significant robustness towards image noise, though our unsupervised training strategy.

\noindent\textbf{JPEG Compression}
In the DSR case, where SR is performed within the same domain, our approach achieves a significantly lower LPIPS distance compared to state-of-the-art. For CSR, where the task is to additionally clean the input from artifacts, our approach has a strong advantage, reducing the LPIPS by more than $0.17$. However, as shown in Figure~\ref{fig:stateOfTheArt_div2k}, the difference in perceptual quality is not fully captured by the quantitative results. All compared approaches produces highly visible block artifacts, stemming from the JPEG compression of the input image. In contrast, our approach provides visually pleasing output without such artifacts.

\begin{figure}[t]	
	\newcommand{\wid}{0.11\linewidth}
	\begin{minipage}{0.22\linewidth}%
		\includegraphics[width=\linewidth]{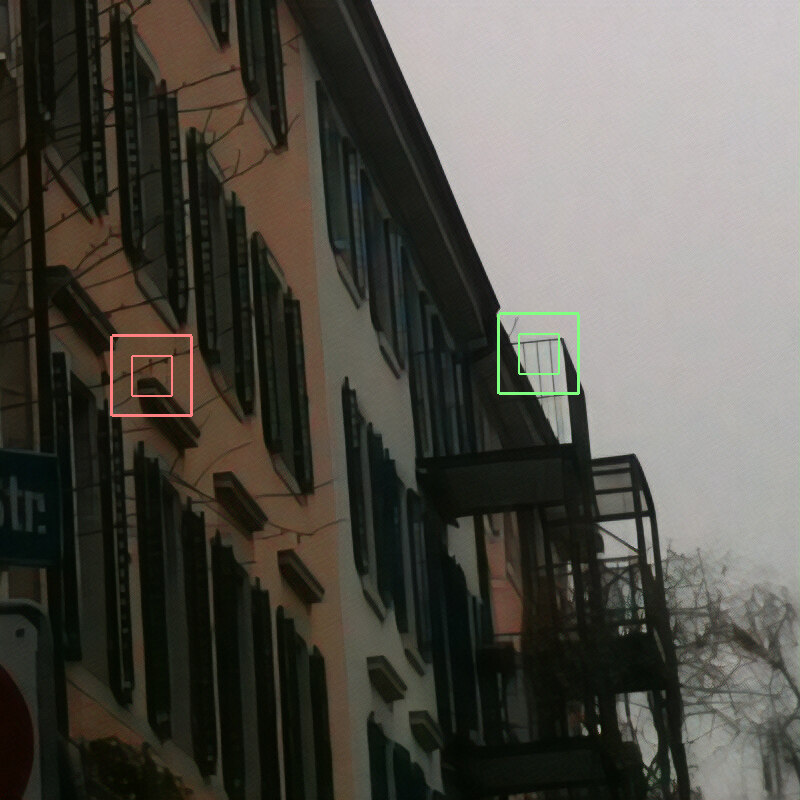}
		\includegraphics[width=\linewidth]{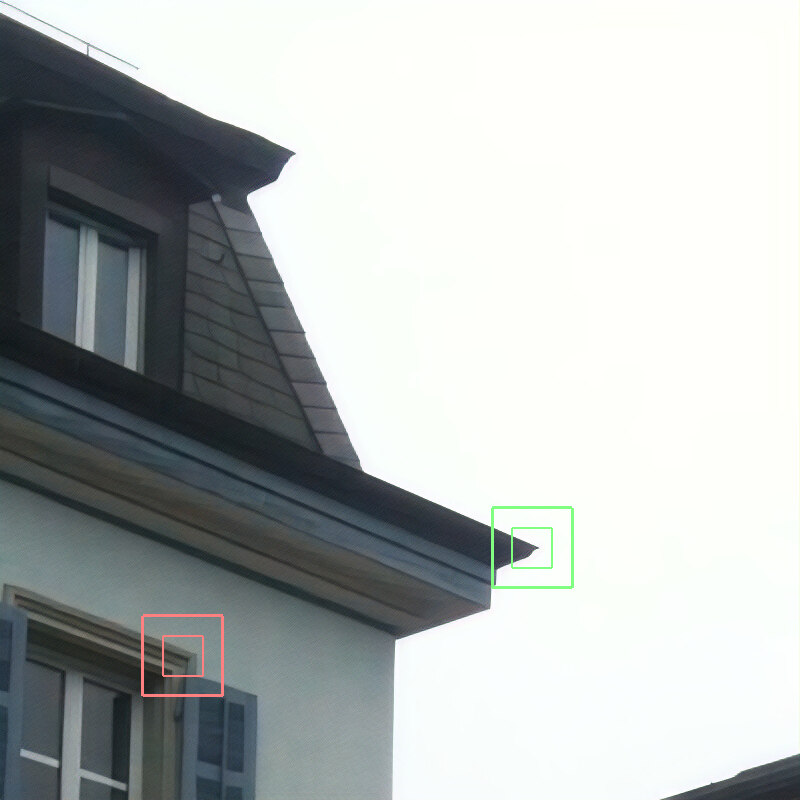}
		\includegraphics[width=\linewidth]{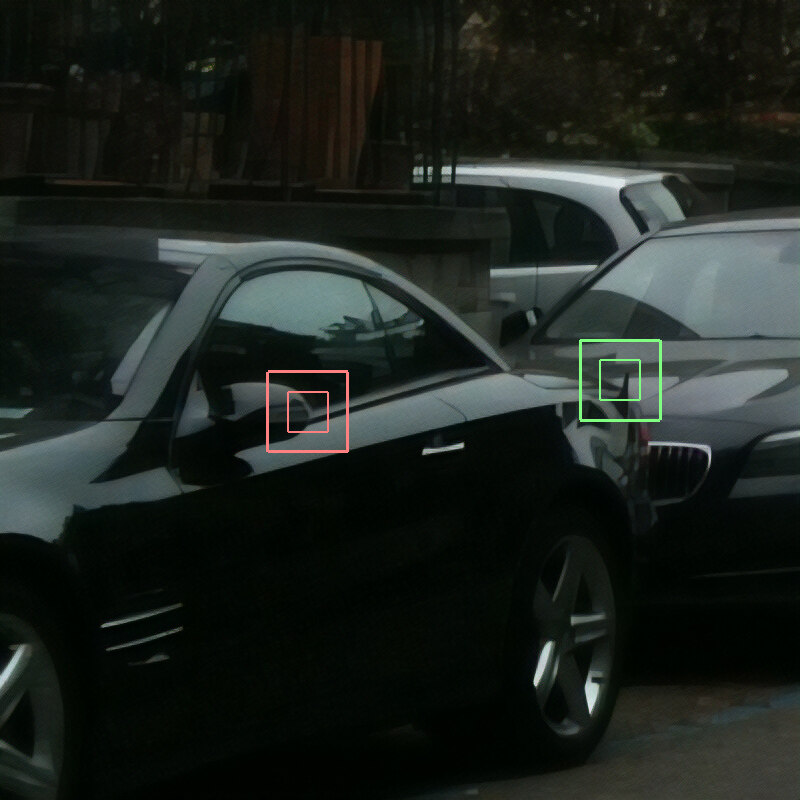}
	\end{minipage}%
	\begin{minipage}{\wid}%
		\includegraphics[width=\linewidth]{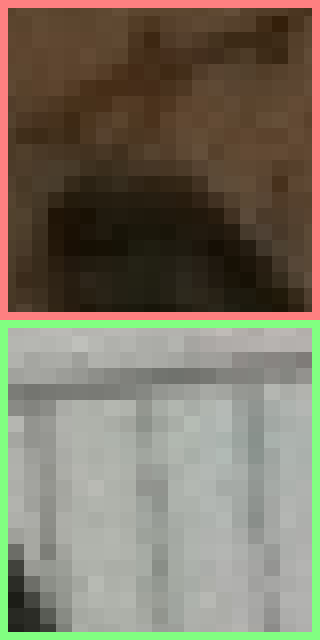}
		\includegraphics[width=\linewidth]{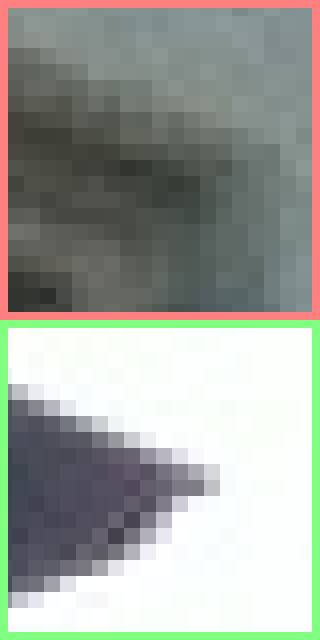}
		\includegraphics[width=\linewidth]{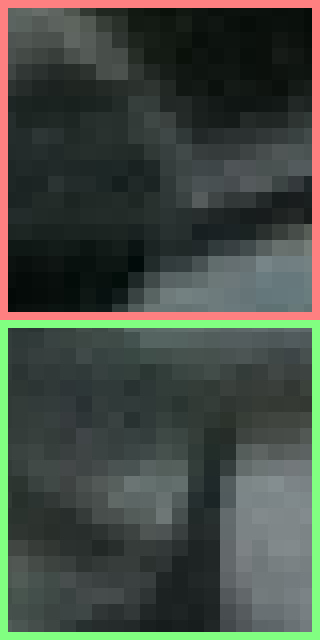}
	\end{minipage}%
	\begin{minipage}{\wid}%
		\includegraphics[width=\linewidth]{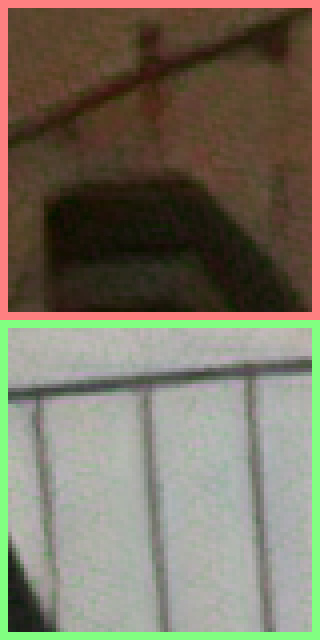}
		\includegraphics[width=\linewidth]{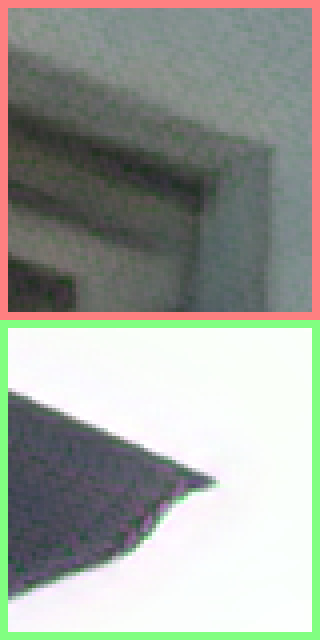}
		\includegraphics[width=\linewidth]{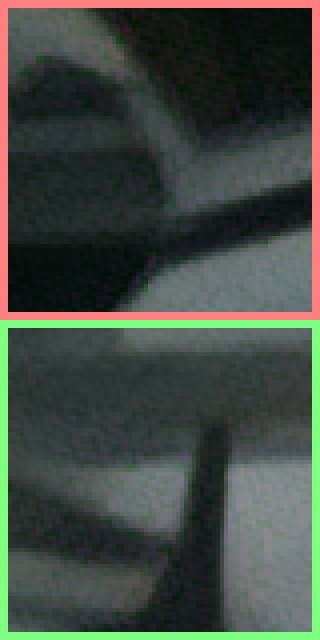}
	\end{minipage}%
	\begin{minipage}{\wid}%
		\includegraphics[width=\linewidth]{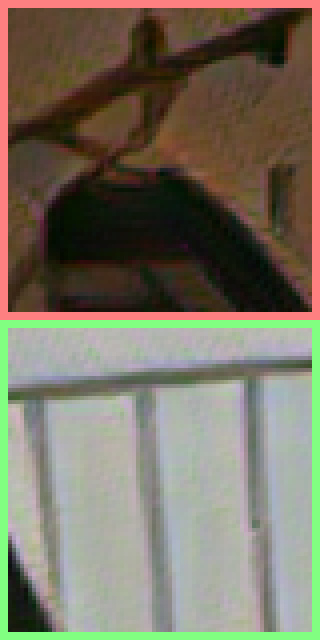}
		\includegraphics[width=\linewidth]{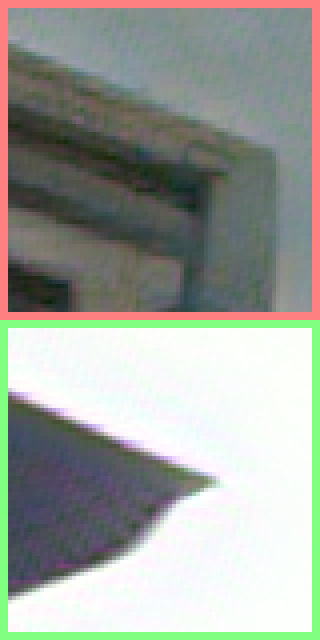}
		\includegraphics[width=\linewidth]{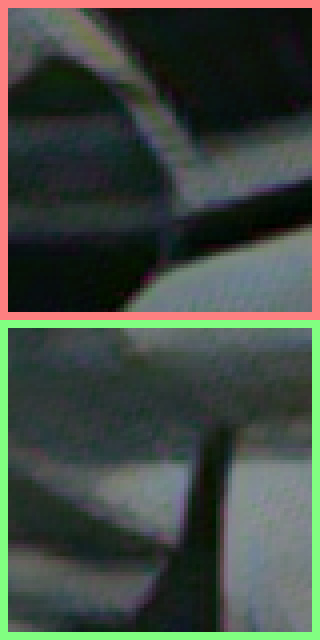}
	\end{minipage}%
	\begin{minipage}{\wid}%
		\includegraphics[width=\linewidth]{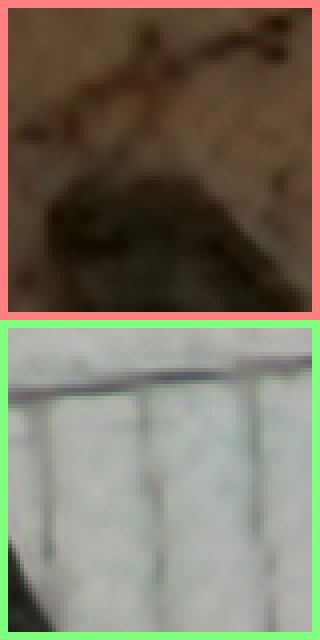}
		\includegraphics[width=\linewidth]{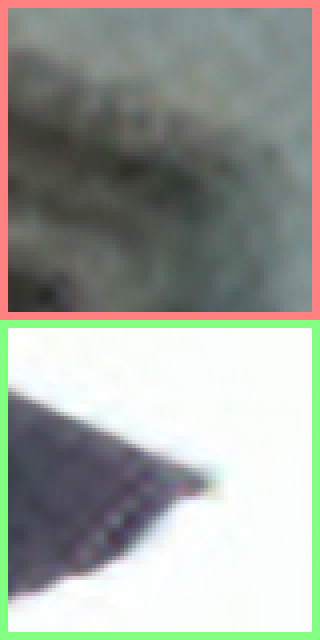}
		\includegraphics[width=\linewidth]{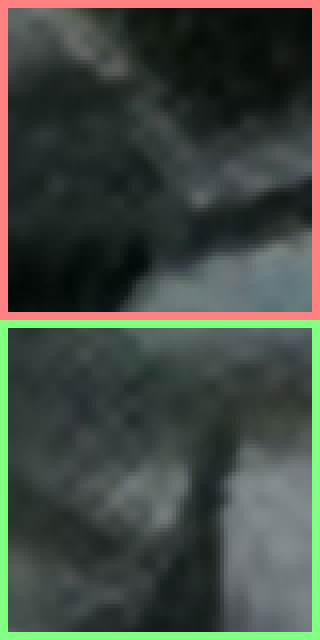}
	\end{minipage}%
	\begin{minipage}{\wid}%
		\includegraphics[width=\linewidth]{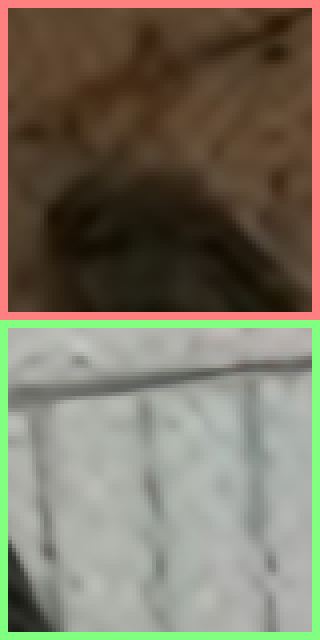}
		\includegraphics[width=\linewidth]{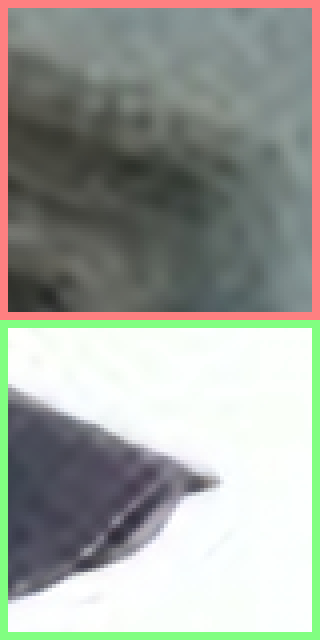}
		\includegraphics[width=\linewidth]{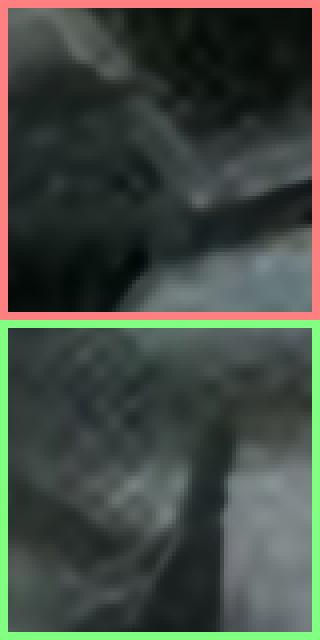}
	\end{minipage}%
	\begin{minipage}{\wid}%
		\includegraphics[width=\linewidth]{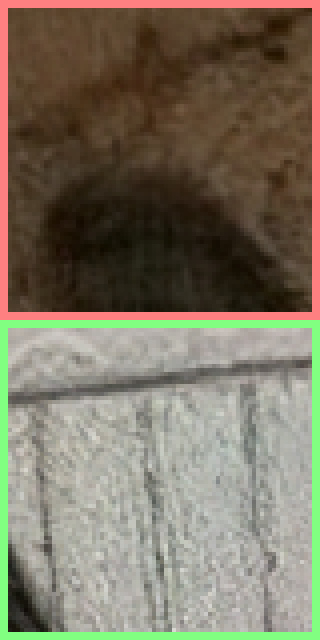}
		\includegraphics[width=\linewidth]{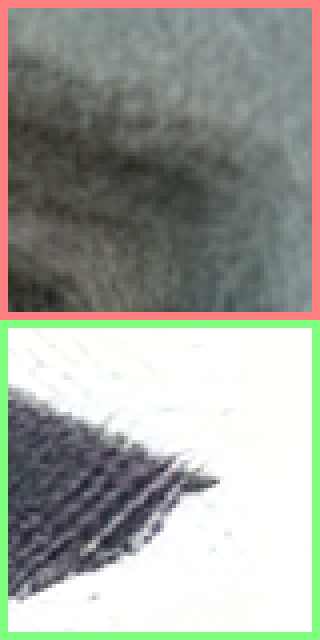}
		\includegraphics[width=\linewidth]{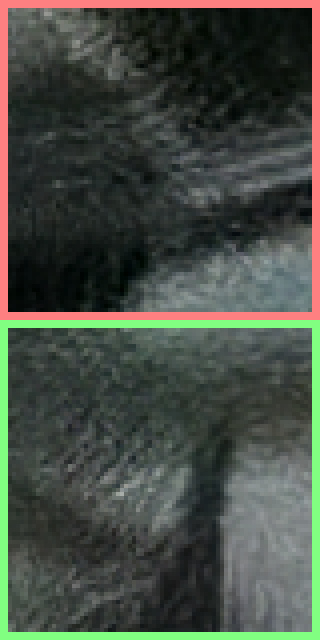}
	\end{minipage}%
	\begin{minipage}{\wid}%
		\includegraphics[width=\linewidth]{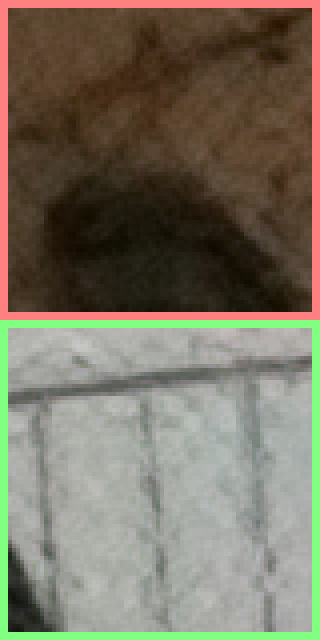}
		\includegraphics[width=\linewidth]{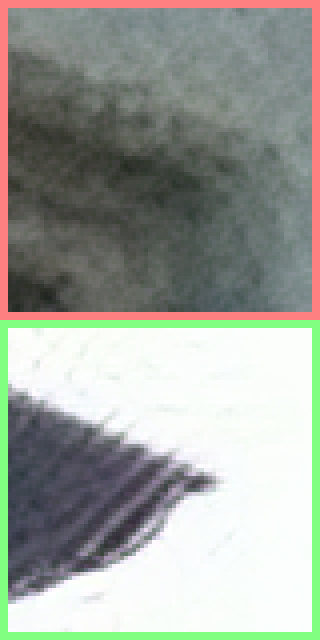}
		\includegraphics[width=\linewidth]{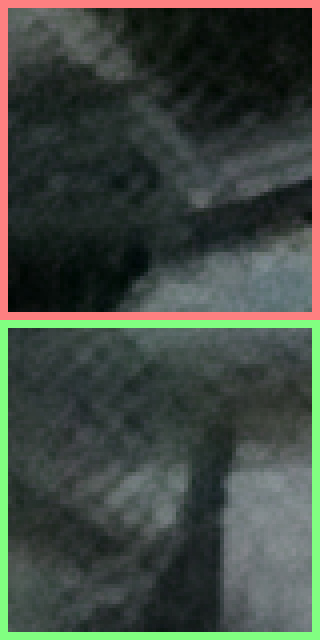}
	\end{minipage}
	\vspace{0mm}
	\newcommand{\sep}{~}
	\resizebox{\linewidth}{!}{
		\begin{tabular}{@{}C{4cm}@{\sep}C{2cm}@{\sep}C{2cm}@{\sep}C{2cm}@{\sep}C{2cm}@{\sep}C{2cm}@{\sep}C{2cm}@{\sep}C{2cm}@{\sep}C{2cm}@{}}
			\textbf{Ours (DSR)}  & Input & \textbf{Ours (DSR)} & \textbf{Ours (CSR)} & ZSSR & EDSR & ESRGAN & ESRGAN FT
		\end{tabular}
	}
	\vspace{0mm}
	\caption{Qualitative comparison on real-world images from the DEPD dataset. Due to the training setup of other state-of-the-art methods they produce large artifacts on real-world data while our methods (DSR, CSR) can super-resolve those images in a perceptual satisfying manner.}
	\label{fig:stateOfTheArt_dped}
\end{figure}

\subsection{Real-World Evaluation on the DPED Dataset}

In this section we apply our method to the original images of the DPED iPhone3 dataset~\cite{ignatov2017dslr}. It contains natural images, which include real-world degradations due to poor sensor and lens quality. We train our model using the training split to represent both the input and output distributions. We also finetune the ESRGAN model on the same data, as described in the previous experiment. Since ground-truth images are not available in this real-world setting, we show a diverse set of qualitative examples from the DPED iPhone3 validation set in Figure~\ref{fig:stateOfTheArt_dped}. The artifacts produced by ZSSR, EDSR, ESRGAN and ESRGAN Finetuned are of similar nature as in the sensor noise case in the DIV2K setting in Figure~\ref{fig:stateOfTheArt_div2k}. Note that these limitations in previous approaches cannot be alleviated by more training data or architectural designs. Instead these issues originate from an oversimplified problem formulation not reflected in most real-world applications. Our approach is able to overcome the limitations of previous methods by learning the input image distribution. Our approach generate high-quality images, with very few artifacts.

\section{Conclusion}
We tackle the problem of real-world super-resolution, where no paired data is available. To avoid the artifacts caused by bicubic downsampling, we learn a network that restores the low resolution image to the real-world image distribution. This allows us to generate realistic training pairs for our super-resolution model. Lastly, we propose a benchmark, based on the DIV2K dataset, for quantitatively evaluating real-world super-resolution approaches. Experiments are performed on our real-world benchmark and the DPED datasets. Compared previous methods, our approach generalizes to natural images, affected by significant sensor noise, compression artifacts and other effects.

\paragraph{Acknowledgements. }
This work was partly supported by ETH General Fund, Amazon through an AWS grant, Nvidia through a GPU grant, and Huawei.


{\small
\bibliographystyle{ieee}
\bibliography{egbib}
}

\clearpage

\setcounter{equation}{0}
\setcounter{figure}{0}
\setcounter{table}{0}
\setcounter{section}{0}

\renewcommand{\theequation}{S\arabic{equation}}
\renewcommand{\thefigure}{S\arabic{figure}}
\renewcommand{\thetable}{S\arabic{table}}
\renewcommand{\thesection}{S\arabic{section}}

\begin{center}
	\textbf{\large Supplementary Material}
\end{center}

In this supplementary material we provide additional details and results. In Sections~\ref{sec:aim} and \ref{sec:ntire} we perform an additional experiment on the AIM 2019 and NTIRE 2018 challenge datasets respectively. Section~\ref{sec:baselines} provides illustrations and further details of the approaches analyzed in the ablative experiments. We provide further details of our real-world SR benchmarking procedure in Section~\ref{sec:benchmark}. Example output images of our learned domain transfer mapping $G$ is shown in Section~\ref{sec:G-visual}. We show additional qualitative visual results in Section~\ref{sec:visual}. Finally, we analyze failure cases in Section~\ref{sec:fail}.

\section{AIM 2019 Real-World Super Resolution Challenge}
\label{sec:aim}
In addition to the experiment settings presented in section~\ref{sse:experimental_setup}, we introduce the real-world SR benchmark employed in the recent AIM 2019 Real-World Super Resolution Challenge \cite{AIM2019RWSRchallenge}. We use the same overall procedure as described in section~\ref{sse:experimental_setup}, but employ a more complex degradation mapping, composed of several different operations. The input domain train images $\{X_i\}_{i=1}^M$ are obtained by directly adding the degradation to the high-resolution Flickr2K~\cite{timofte2017ntire} images, while we use the clean DIV2K~\cite{div2k} train split for the target domain set $\{Y_j\}_{j=1}^N$. For validation and testing we use the corresponding splits from the DIV2K. The results of our approach compared to state-of-the-art methods are shown in Table~\ref{tab:stateOfTheArtAIMVa}~and~\ref{tab:stateOfTheArtAIMTe}. Note that Track~1 and 2 correspond to the DSR and CRS settings respectively. Our approach achieves superior LPIPS score compared to previous approaches.

\begin{table}[bp]
	\centering
	\newcommand{\sep}{~~}
	\resizebox{0.7\columnwidth}{!}
	{
		\begin{tabular}{@{}ll@{\sep}|@{\sep}c@{\sep}c@{\sep}c@{\sep}c@{}}
			& Method        & $\uparrow$PSNR & $\uparrow$SSIM & $\downarrow$LPIPS \\
			\hline
			\multirow{5}{1mm}{\rotatebox{90}{\resizebox{7mm}{!}{DSR}}}
            & ZSSR          & 25.24 & 0.61 & 0.6613 \\
            & EDSR          & 25.24 & 0.60 & 0.6698 \\
            & ESRGAN        & 22.49 & 0.50 & 0.5829 \\
            & ESRGAN FT     & 24.60 & 0.55 & 0.4482 \\
            & \textbf{Ours} & 24.81 & 0.56 & 0.4376 \\
			\hline
			\multirow{5}{1mm}{\rotatebox{90}{\resizebox{7mm}{!}{CSR}}}
            & ZSSR          & 22.42 & 0.61 & 0.5996 \\
            & EDSR          & 22.36 & 0.60 & 0.6150 \\
            & ESRGAN        & 20.69 & 0.51 & 0.5604 \\
            & ESRGAN FT IN  & 21.40 & 0.52 & 0.5191 \\
            & ESRGAN FT OUT & 21.66 & 0.55 & 0.5282 \\
            & \textbf{Ours} & 21.59 & 0.55 & 0.4720 \\
		\end{tabular}
	}
	\vspace*{0mm}
	\caption{Comparison to state-of-the-art super-resolution methods on the AIM Real World Super-Resolution Challenge dataset using the DSR (Track 1) and CSR (Track 2) settings on the validation set.}
	\label{tab:stateOfTheArtAIMVa}
	\vspace*{-4mm}
\end{table}

\begin{table}[bp]
	\centering
	\newcommand{\sep}{~~}
	\resizebox{0.7\columnwidth}{!}
	{
		\begin{tabular}{@{}ll@{\sep}|@{\sep}c@{\sep}c@{\sep}c@{\sep}c@{}}
			& Method        & $\uparrow$PSNR & $\uparrow$SSIM & $\downarrow$LPIPS \\
			\hline
			\multirow{7}{1mm}{\rotatebox{90}{\resizebox{7mm}{!}{DSR}}}
            & Bicubic          &  25.34 &  0.61 &  0.7331 \\
            & EDSR             &  25.14 &  0.60 &  0.6657 \\
            & ESRGAN           &  22.57 &  0.51 &  0.5770 \\
            & ESRGAN FT        &  24.54 &  0.56 &  0.4458 \\
            & \textbf{Ours}    &  24.22 &  0.54 &  0.4356 \\
            \hline
            & Fully Supervised &  24.22 &  0.55 &  0.3011 \\
            \hline
            \hline
            \multirow{8}{1mm}{\rotatebox{90}{\resizebox{7mm}{!}{CSR}}}
            & Bicubic       &  22.37 &  0.63 &  0.6602 \\
            & EDSR          &  22.35 &  0.62 &  0.5959 \\
            & ESRGAN        &  20.76 &  0.52 &  0.5539 \\
            & ESRGAN FT IN  &  21.36 &  0.54 &  0.5275 \\
            & ESRGAN FT OUT &  21.94 &  0.59 &  0.5083 \\
            & \textbf{Ours} &  21.17 &  0.54 &  0.4594 \\
            \hline
            & Fully Supervised &  22.80 &  0.65 &  0.2928 \\
		\end{tabular}
	}
	\vspace*{0mm}
	\caption{Comparison to state-of-the-art super-resolution methods on the AIM Real World Super-Resolution Challenge dataset using the DSR (Track 1) and CSR (Track 2) settings on the test set.}
	\label{tab:stateOfTheArtAIMTe}
	\vspace*{-4mm}
\end{table}

\begin{table*}[bp]
	\centering
	\resizebox{0.8\linewidth}{!}
	{
		\begin{tabular}{lcccccccccc}
			\toprule
			& Bicubic & ZSSR & EDSR & ESRGAN (Baseline) & Cleaning the Input & Low resolution supervision & \textbf{Ours} \\
			\midrule
			LPIPS$\downarrow$ & 0.2537         & 0.2320 & 0.2534 & 0.2670 & 0.2272 & 0.2038 & \textbf{0.1858} \\
			PSNR$\uparrow$    & \textbf{22.72} & 22.66  & 22.48  & 21.10  & 20.56  & 19.98  & 18.89           \\
			SSIM$\uparrow$    & \textbf{0.52}  & 0.51   & 0.49   & 0.37   & 0.42   & 0.47   & 0.49            \\
			\bottomrule
		\end{tabular}
	}\vspace{1mm}
	\caption{
		Comparison of the different state-of-the-art methods and three versions of our ablation study for the DSR. We validate the methods on the DIV2K Track 2 dataset, which was published for the NTIRE2018 Challenge. Our  approach  achieves  superior perceptual quality measured by the LPIPS distance. }
	\label{tab:X4Degraded}
\end{table*}

\section{NTIRE 2018 Evaluation}
\label{sec:ntire}

We perform an additional experiment on the NTIRE 2018 \cite{timofte2018ntire} challenge. To evaluate the approaches in the unsupervised scenario, we use the \textbf{Track 2: Realistic Mild $\times 4$ adverse conditions}. In this setting, the an unknown degradation operation has been applied to the low resolution images to simulate realistic sensor noise and motion blur. The goal is to produce clean output images, similar to our CSR scenario. Although LR and HR training pairs are provided, we do not utilize the paired data. As performed in the paper, we train our approach in the fully unsupervised real-world setting, using only \emph{unpaired} LR and HR images. Compared to the official challenge \cite{timofte2018ntire}, our approach is thus trained in considerably harder, but more realistic conditions.

The results are provided in Table~\ref{tab:X4Degraded}. We compare our approach to simple bicubic upscaling, ZSSR \cite{shocher2018zero}, EDSR \cite{lim2017EDSR} and the approaches considered in our ablative study: the baseline ESRGAN \cite{wang2018esrgan}, Cleaning the Input, and Low resolution supervision. See Section~4.2 in the main paper and Section~\ref{sec:baselines} here for more details about the latter baseline versions. Results are reported in terms of LPIPS \cite{zhang2018unreasonable}, PSNR and SSIM. Since our goal is perceptual super resolution, we focus on the LPIPS metric. A comparison between these metrics are shown in Figure~\ref{fig:psnr_vs_lpips}. Although na\"ive bicubic upscaling achieves the best PSNR in Table~\ref{tab:X4Degraded}, our approach clearly achieves better perceptual results, as indicated by the LPIPS metric. Among the compared previous SR methods, the baseline ESRGAN achieves the best performance with an LPIPS of $0.2534$. Employing the generator $F$ to clean the input image during inference improves the LPIPS metric to $0.2272$. Using supervision in the low resolution domain yields an LPIPS of $0.2038$. Our approach, based on the domain distribution learning, allowing direct supervision in the high resolution domain, achieves the best LPIPS of $0.1858$. This additional experiment demonstrates the same trend as the results shown in the main paper.

\begin{figure*}[t]
	\centering
	\newcommand{\wid}{0.4\textwidth}%
	\newcommand{\scl}{0.5}%
	\subfloat[Baseline\label{fig:ablation_baseline}]{
		\includegraphics[scale=\scl]{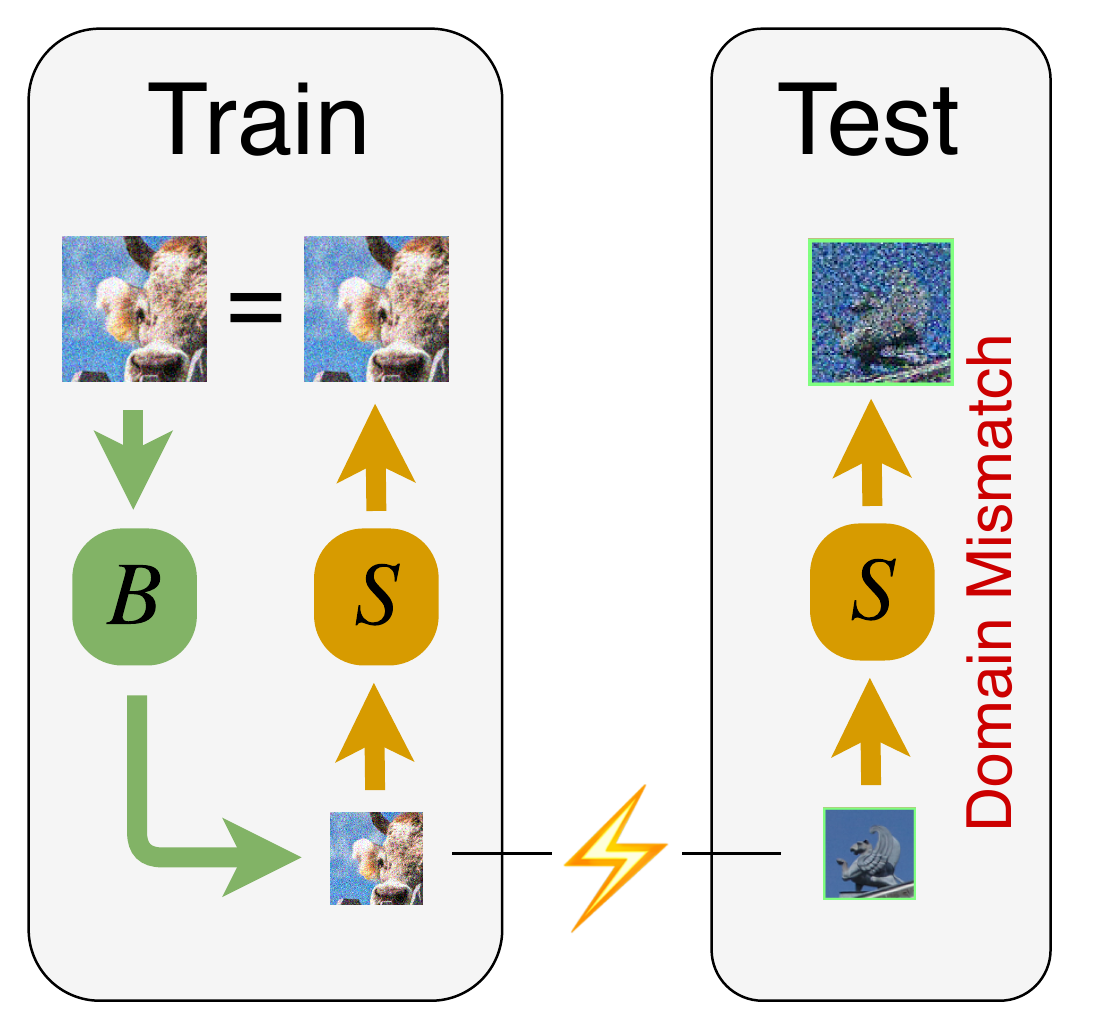}}\hspace{15mm}
	\subfloat[Cleaning the input\label{fig:ablation_clean}]{
		\includegraphics[scale=\scl]{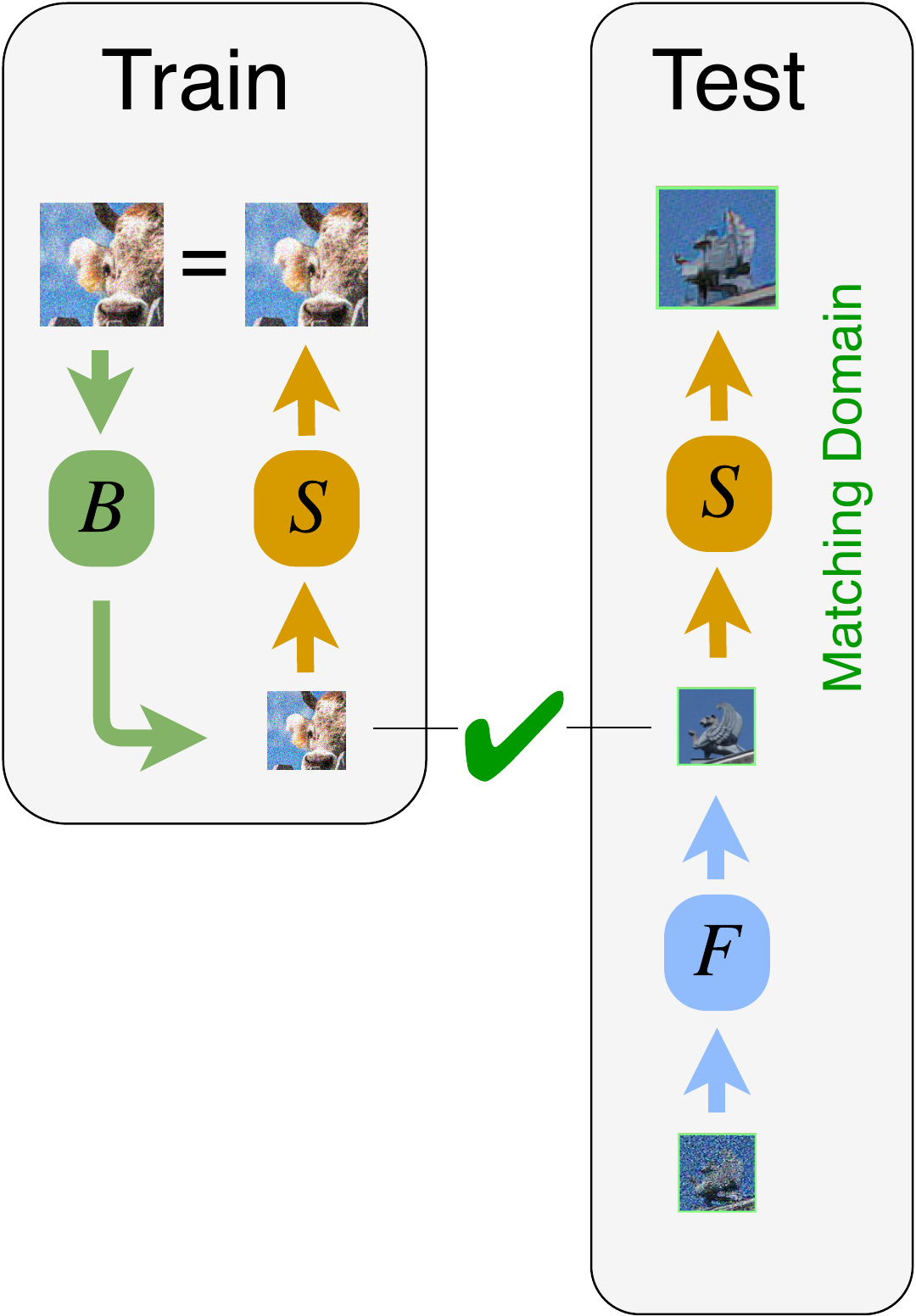}}\vspace{5mm}
	
	\subfloat[Low resolution supervision\label{fig:ablation_lr}]{
		\includegraphics[scale=\scl]{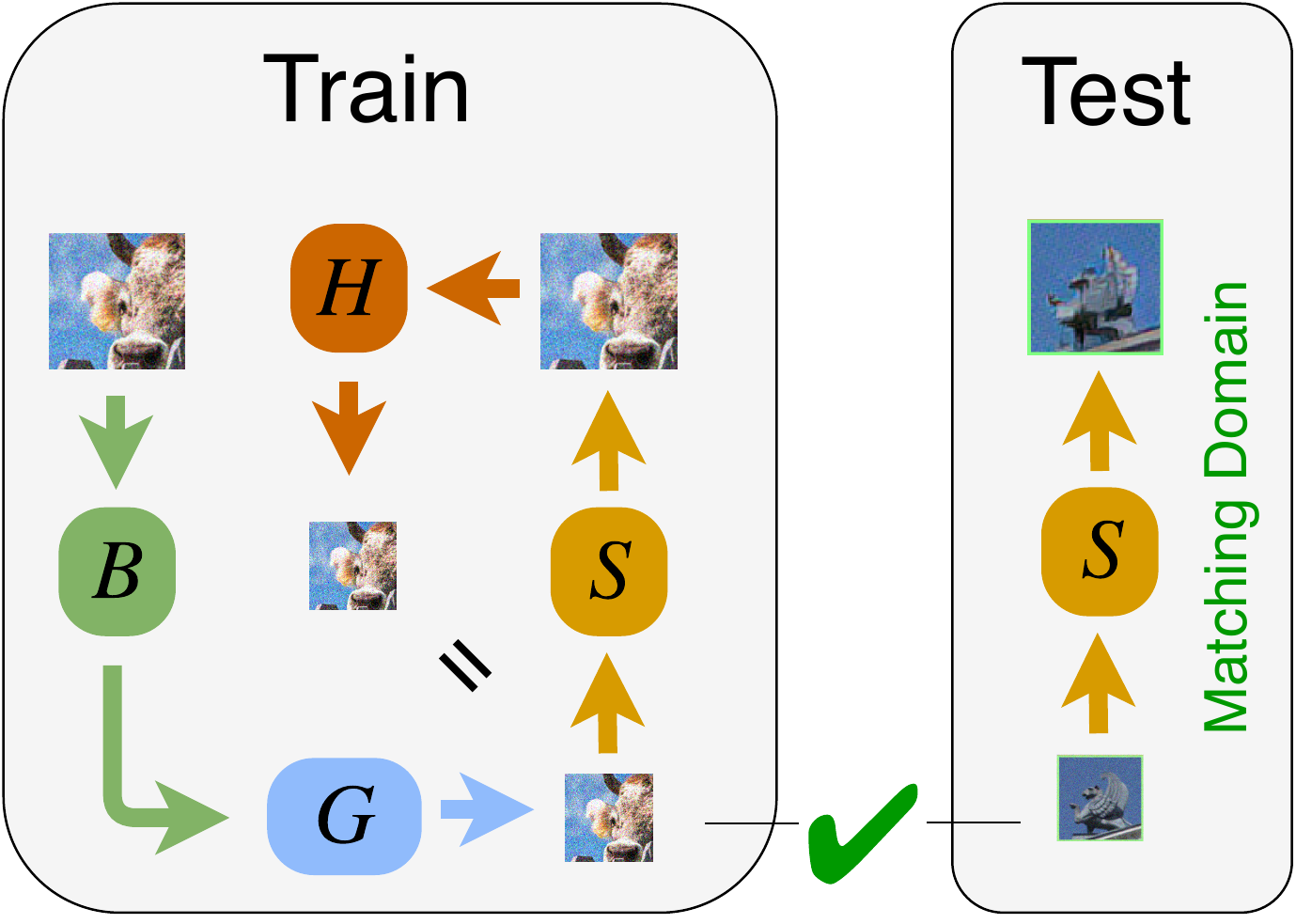}}\hspace{15mm}
	\subfloat[Ours\label{fig:ablation_ours}]{
		\includegraphics[scale=\scl]{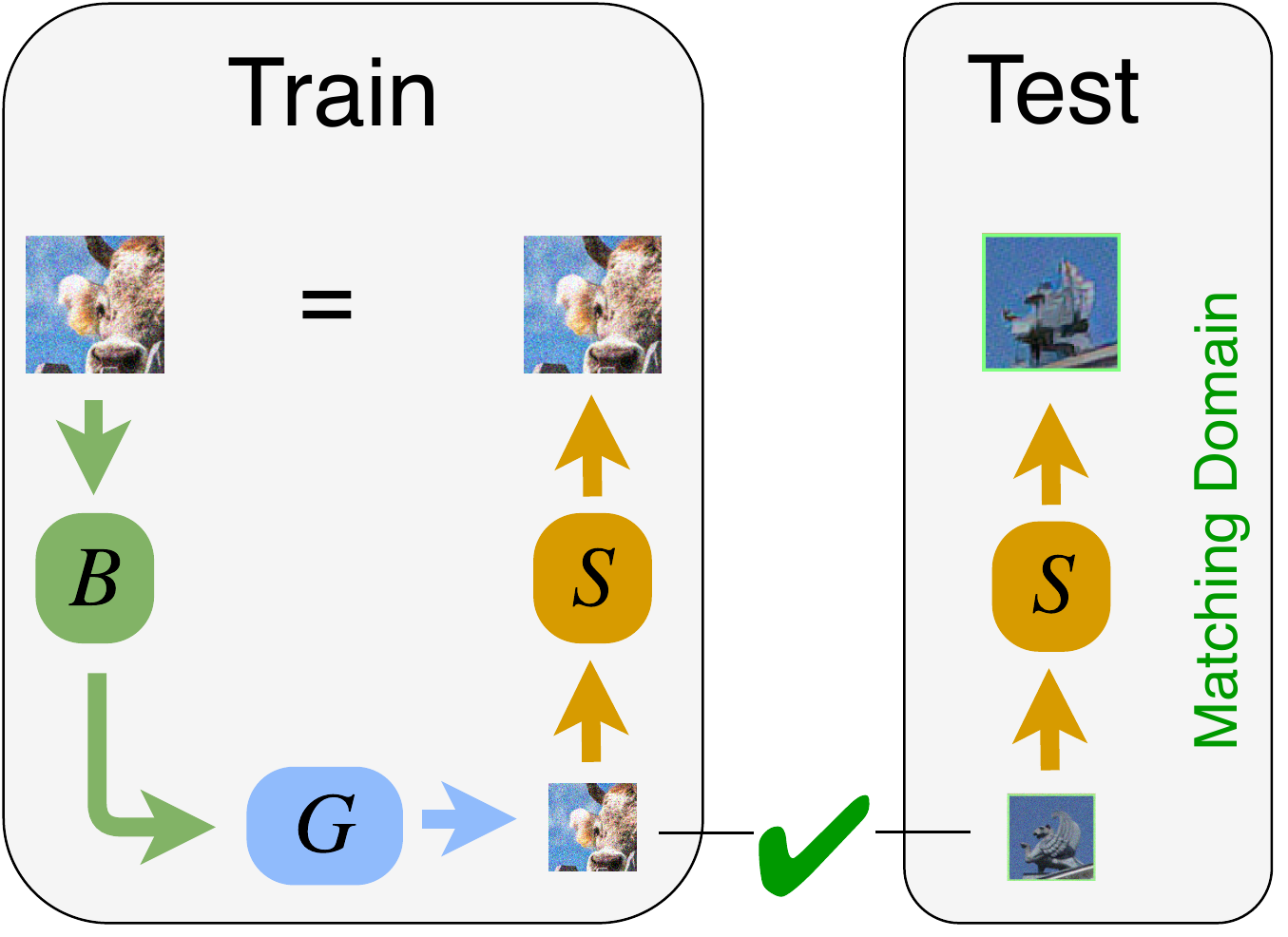}}\vspace{5mm}
	
	\includegraphics[width=\textwidth]{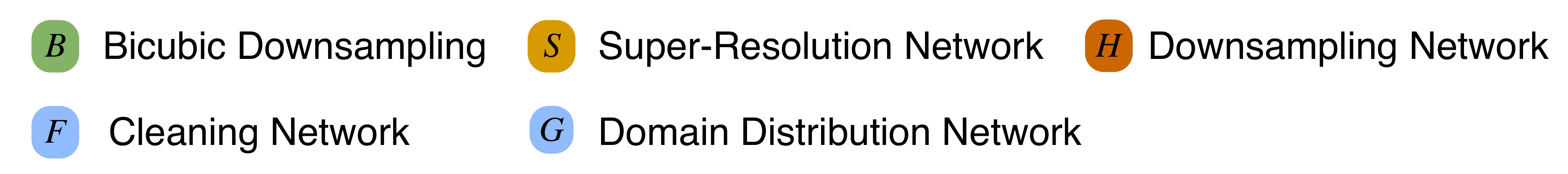}\vspace{5mm}
	\caption{The four different approaches for real-world super-resolution analyzed in the ablation study (Section~4.2 in the paper). For each method, we illustrate how the super-resolution network $S$ is trained (left) and applied during inference. Pixel-wise supervision is indicated by ``$=$'', while the discriminator losses are omitted for clarity.}
	\label{fig:ablationarc}
\end{figure*}

\section{Details of the Baseline Approaches}
\label{sec:baselines}

Here we provide additional details of the ablative methods evaluated in Section 4.2 in the paper. Figure~\ref{fig:ablationarc} illustrate the three baseline versions, along with our approach. In the baseline ESRGAN version (Figure~\ref{fig:ablation_baseline}), we follow the same protocol to generate the low resolution training samples as ESRGAN to finetune the pretrained model. The super-resolution network is trained to reconstruct images that are bicubic downsampled by $B$. As shown, this produces strong artifacts due to the mismatch of input domains during the training and testing phases. A first approach to match the input distribution of the super-resolution network is to first clean the low resolution image during test time (Figure~\ref{fig:ablation_clean}). The domain distribution mapping $F$ is learned using cycle consistency loss to map input images to the domain of bicubic downsampled images. Another way to obtain matching domains during train and test time is to do supervision in low resolution, as one can directly use the natural image as low resolution input for training (Figure~\ref{fig:ablation_lr}).

In Our approach (Figure~\ref{fig:ablation_ours} we make sure that the inputs during train and test are matching by applying the mapping $G$ during training to invert the effects of bicubic downsampling on the image distribution. This mapping is learned by our with unsupervised domain distribution learning (Section~3.3). Compared to the version in Figure~\ref{fig:ablation_clean}, our approach requires no extra network for processing the input image during test-time, before the super-resoution is applied. Instead, our approach leans to super-resolve natural images directly. Moreover, in contrast to Figure~\ref{fig:ablation_lr}, we do a pixel-wise supervision in high resolution with the original natural image, which helps the super-resolution network to produce crisp, photo-realistic images.

\section{Real-world SR Benchmarking Details}
\label{sec:benchmark}

Here, we provide additional details of the proposed benchmarking procedure and dataset for real-world super-resolution methods. 
In classical super-resolution, the low resolution images for training and evaluation are constructed by applying bicubic downsampling to a dataset of natural images. Therefore, those methods are only evaluated on how well they super-resolve images that were downsampled the same way as during training. However, in the real-world setting the aim is to super-resolve natural images to a higher resolution that is unseen during training. Thus, the desired high resolution images are not available for training the super-resolution network. To reflect this crucial aspect in the proposed benchmarking procedure for real-world super-resolution, we apply different procedures for constructing training and test data. Importantly, the training data only contains low resolution images, while the original high resolution images are only used when evaluating the methods. Thus the final desired image resolution is unseen during training, as in real-world applications.

We construct the train and test images for real-world super-resolution, based on a dataset of original images. An illustration of the procedure is shown in Figure~\ref{fig:data-pipeline} (also shown in the main paper). For evaluation we construct an input-output image pair. The input image is obtained by first downsampling the original image and then applying the degradation transformation to simulate the real-world case. Since the degradation is heavily affected by image downsampling, it is always applied after the downsampling procedure. In the domain-specific super-resolution case (DSR), where the goal is to super-resolve the image while preserving the original characteristics, we construct the ground-truth output image by applying the same degradation operation directly to the original image. In the clean super-resolution case, we employ the original image as ground-truth.
The training set is constructed by first downsampling the original image and then applying the degradation operation. In the CSR case, a set of clean images are also available during training (see Section~4.1 in the paper). These are constructed by only applying bicubic downsampling to the original images, and no degradations. Note that no images of the original resolution, used for evaluation, are available for training.

\begin{table}[b]
	\centering
	\begin{tabular}{l|ccc}
		Dataset & Num.~images & Mean image size & Set \\
		\toprule
		$\text{DF2K}$ & 3450 & 1439x1935 & Training \\
		$\text{DIV2K}$ & 100 & 1451x1979 & Testing \\
	\end{tabular}\vspace{1mm}
	\caption{Datasets that were used for our experiments.}
	\label{tab:Datasets}
\end{table}

In this work, we employ the DF2K \cite{wang2018esrgan} dataset, consisting of DIV2K \cite{div2k} and Flikr2K \cite{wang2018esrgan} datasets. The dataset specifications are summarized in Table~\ref{tab:Datasets}. For testing, the 100 validation images in DIV2K are used. The training sets are constructed using the 3450 training images in DF2K. The images for training and evaluation has no overlap. We always employ $4 \times$ downsampling, and evaluate the methods for sensor noise and JPEG degradation, as described in the paper (Section~4.1).  

\section{Domain Mapping Examples}
\label{sec:G-visual}
\begin{figure}[t]	
	\includegraphics[width=0.32\linewidth]{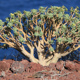}
	\includegraphics[width=0.32\linewidth]{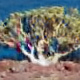}
	\includegraphics[width=0.32\linewidth]{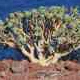}
	
	\includegraphics[width=0.32\linewidth]{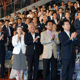}
	\includegraphics[width=0.32\linewidth]{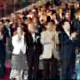}
	\includegraphics[width=0.32\linewidth]{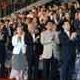}
	
	\includegraphics[width=0.32\linewidth]{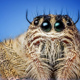}
	\includegraphics[width=0.32\linewidth]{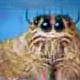}
	\includegraphics[width=0.32\linewidth]{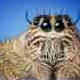}
	
	\includegraphics[width=0.32\linewidth]{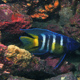}
	\includegraphics[width=0.32\linewidth]{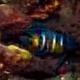}
	\includegraphics[width=0.32\linewidth]{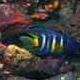}
	
	\includegraphics[width=0.32\linewidth]{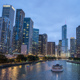}
	\includegraphics[width=0.32\linewidth]{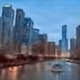}
	\includegraphics[width=0.32\linewidth]{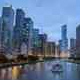}
	
	\resizebox{\linewidth}{!}{
		\begin{tabular}{C{2cm} C{2cm} C{2cm}}
			Input $Z$ & Output $\hat{X}$ & GT $X$
		\end{tabular}
	}
	\vspace{-2mm}
	\caption{Visual example output images of the domain distribution learning network $G$ when trained with images corrupted with \textbf{JPEG compression}. The Input is the bicubic downsampled image $Z$, Output the image that is transferred to the natural distribution $\hat{X} = G(Z)$ and GT show the real degradation applied to $Z$.}
	\label{fig:cycle_jpg}
\end{figure}

As described in Section~3.3 of the main paper, we learn a generator $G$ that transfers the downsampled images to the natural domain. In Figure~\ref{fig:cycle_noise} and Figure~\ref{fig:cycle_jpg} we show how the downsampled images are transferred to images with sensor noise and JPEG artifacts respectively. In the sensor noise case one can see a Gaussian noise like image characteristics. In the JPEG case, the generator $G$ has learned to simulate blocky compression artifacts, especially at sharp edges.

\section{Visual Results}
\label{sec:visual}
\begin{figure*}[t]
	\includegraphics[width=\linewidth]{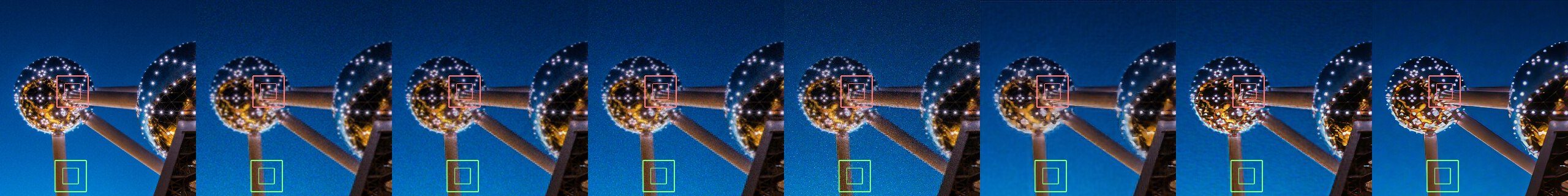}
	\includegraphics[width=\linewidth]{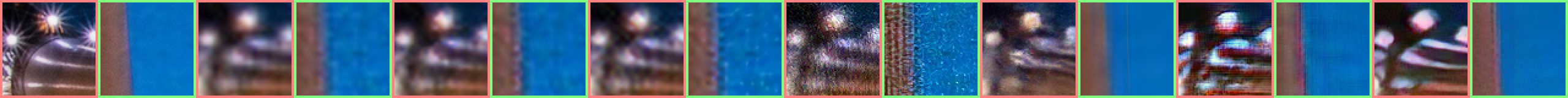}
	\includegraphics[width=\linewidth]{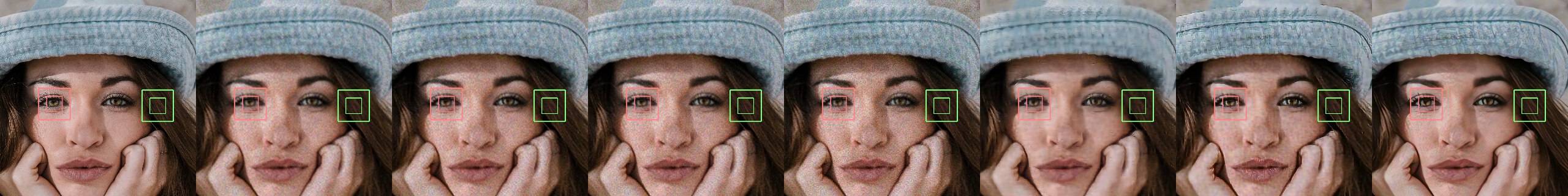}
	\includegraphics[width=\linewidth]{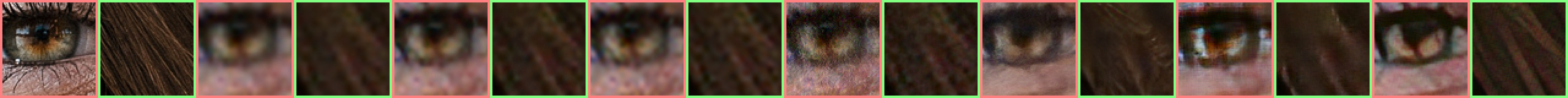}
	\vspace{0mm}
	\newcommand{\sep}{~}
	\resizebox{\linewidth}{!}{
		\begin{tabular}{@{}C{3cm}@{\sep}C{3cm}@{\sep}C{3cm}@{\sep}C{3cm}@{\sep}C{3cm}@{\sep}C{3cm}@{\sep}C{3cm}@{\sep}C{3cm}@{}}
			GT & Bicubic & ZSSR & EDSR & ESRGAN & Cleaning the input & Low resolution supervision & Ours
		\end{tabular}
	}
	\vspace{0mm}
	\caption{Qualitative comparison on the NTIRE2018 Track 2 challenge.}
	\label{fig:div2kr}
\end{figure*}

Visual examples from the NTIRE2018 challenge evaluation (Section~\ref{sec:ntire}) is shown in Figure~\ref{fig:div2kr}. We also show additional visual examples for the real-world DPED iPhone dataset in Figure~\ref{fig:stateOfTheArt_dped}.

\section{Failure Cases}
\label{sec:fail}
\begin{figure}[t]	
	
	\includegraphics[width=0.32\linewidth]{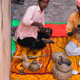}
	\includegraphics[width=0.32\linewidth]{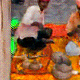}
	\includegraphics[width=0.32\linewidth]{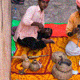}
	
	\includegraphics[width=0.32\linewidth]{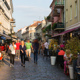}
	\includegraphics[width=0.32\linewidth]{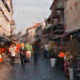}
	\includegraphics[width=0.32\linewidth]{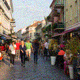}
	
	\includegraphics[width=0.32\linewidth]{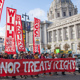}
	\includegraphics[width=0.32\linewidth]{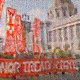}
	\includegraphics[width=0.32\linewidth]{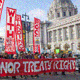}
	
	\includegraphics[width=0.32\linewidth]{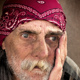}
	\includegraphics[width=0.32\linewidth]{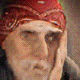}
	\includegraphics[width=0.32\linewidth]{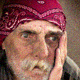}

	\includegraphics[width=0.32\linewidth]{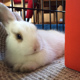}
	\includegraphics[width=0.32\linewidth]{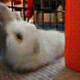}
	\includegraphics[width=0.32\linewidth]{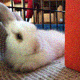}
	
	\resizebox{\linewidth}{!}{
		\begin{tabular}{C{2cm} C{2cm} C{2cm}}
			Input $Z$ & Output $\hat{X}$ & GT $X$
		\end{tabular}
	}
	\vspace{-2mm}
	\caption{Visual example output images of the domain distribution learning network $G$ when trained with images corrupted with \textbf{sensor noise}. The Input is the bicubic downsampled image $Z$, Output the image that is transferred to the natural distribution $\hat{X} = G(Z)$ and GT show the real degradation applied to $Z$.}
	\label{fig:cycle_noise}
\end{figure}

We developed our method in order to match the natural input domain during test time also for real-world datasets with significant noise or JPEG artifacts. This requires a domain distribution learning that also introduces some artifacts by itself as shown in the first row of Figure~\ref{fig:cycle_noise}. When training the network on clean high quality data, for which bicubic downsampling does not have a big effect on the image distribution, our method has slightly more artifacts due to the domain transfer learning.

\clearpage

\begin{figure*}[t]	

	\newcommand{\wid}{0.125\linewidth}
	\begin{minipage}{0.25\linewidth}%
		\includegraphics[width=\linewidth]{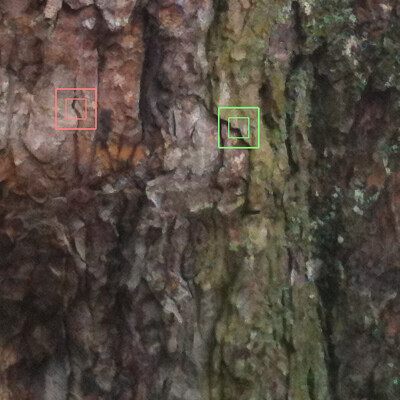}
		\includegraphics[width=\linewidth]{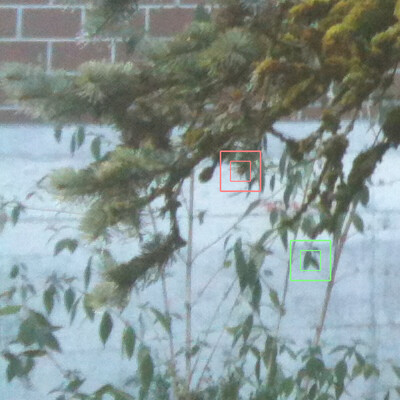}
		\includegraphics[width=\linewidth]{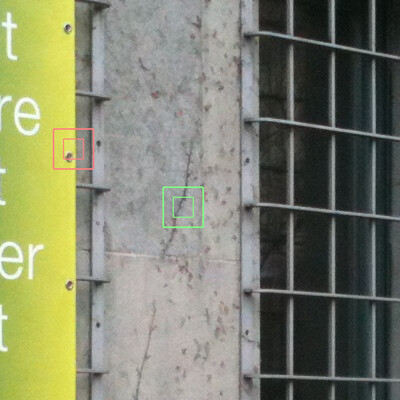}
	\end{minipage}%
	\begin{minipage}{\wid}%
		\includegraphics[width=\linewidth]{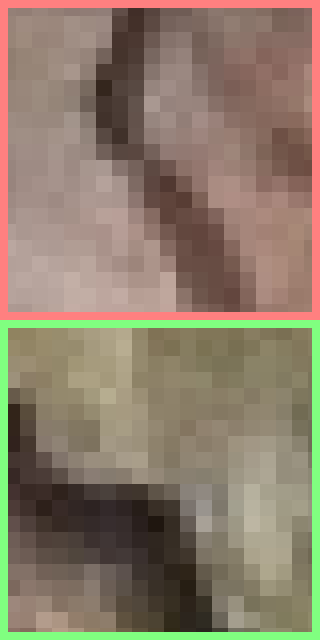}
		\includegraphics[width=\linewidth]{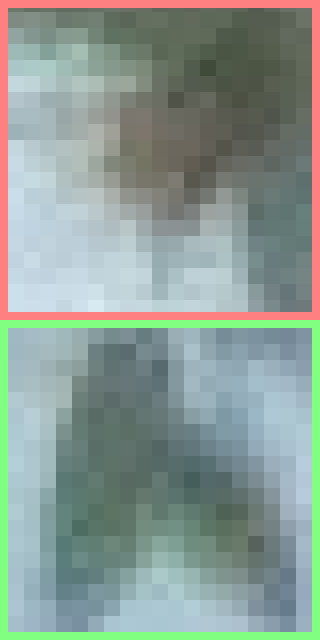}
		\includegraphics[width=\linewidth]{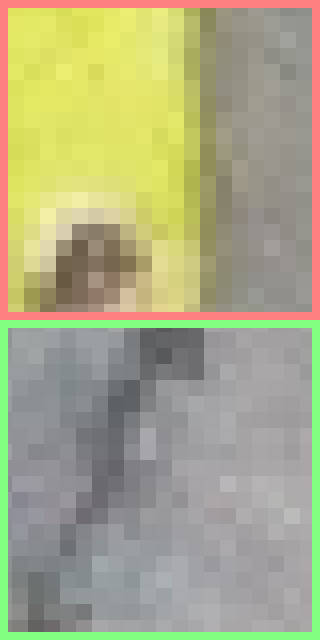}
	\end{minipage}%
	\begin{minipage}{\wid}%
		\includegraphics[width=\linewidth]{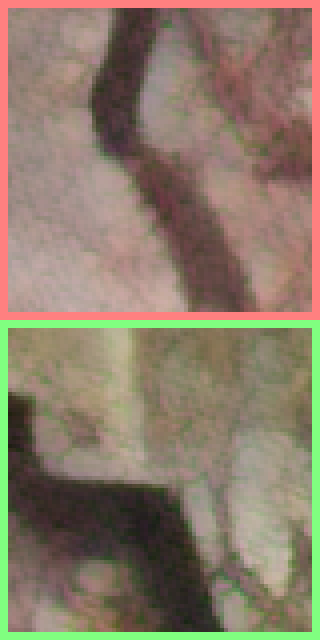}
		\includegraphics[width=\linewidth]{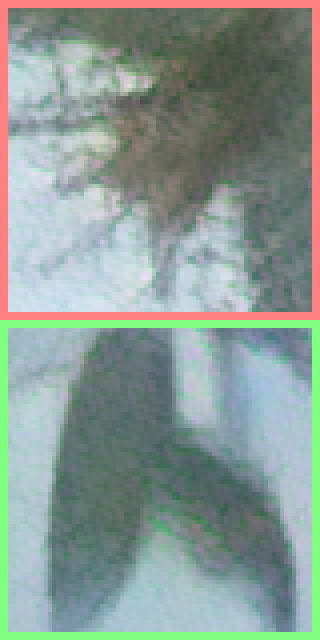}
		\includegraphics[width=\linewidth]{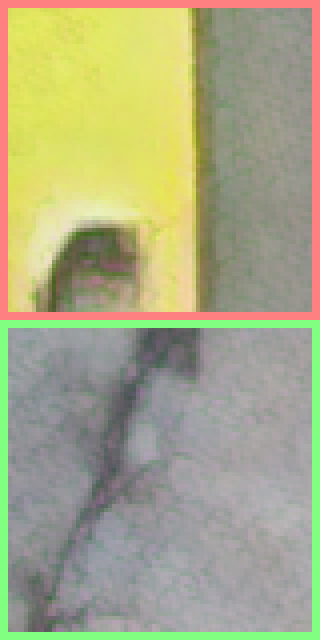}
	\end{minipage}%
	\begin{minipage}{\wid}%
		\includegraphics[width=\linewidth]{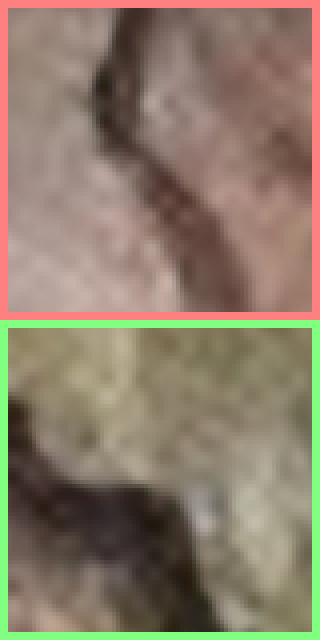}
		\includegraphics[width=\linewidth]{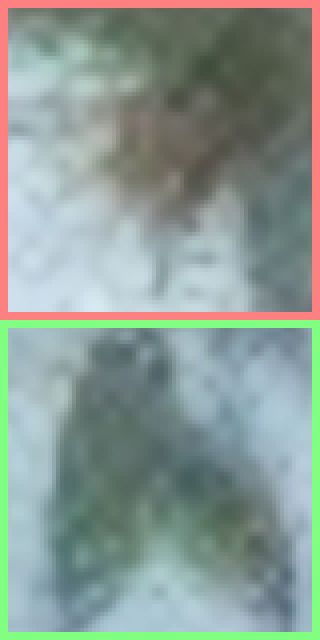}
		\includegraphics[width=\linewidth]{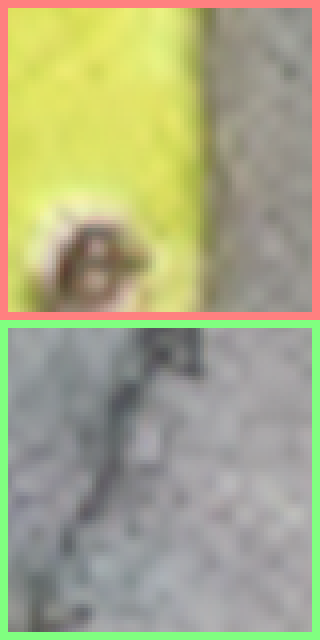}
	\end{minipage}%
	\begin{minipage}{\wid}%
		\includegraphics[width=\linewidth]{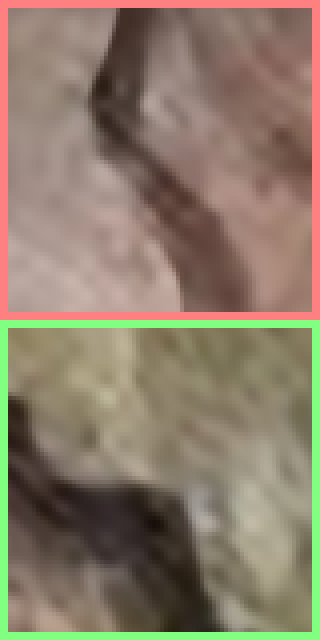}
		\includegraphics[width=\linewidth]{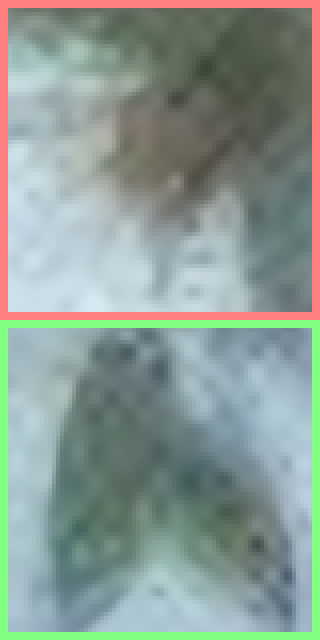}
		\includegraphics[width=\linewidth]{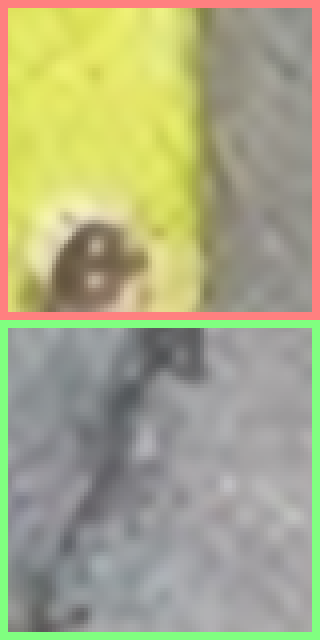}
	\end{minipage}%
	\begin{minipage}{\wid}%
		\includegraphics[width=\linewidth]{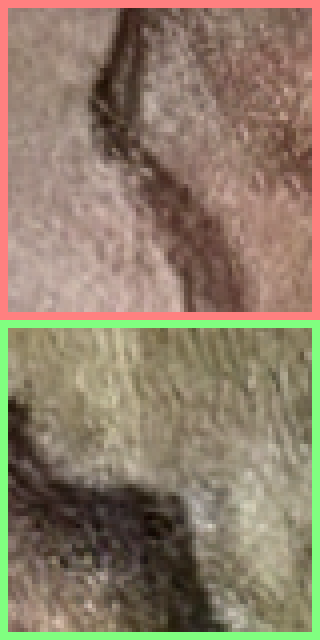}
		\includegraphics[width=\linewidth]{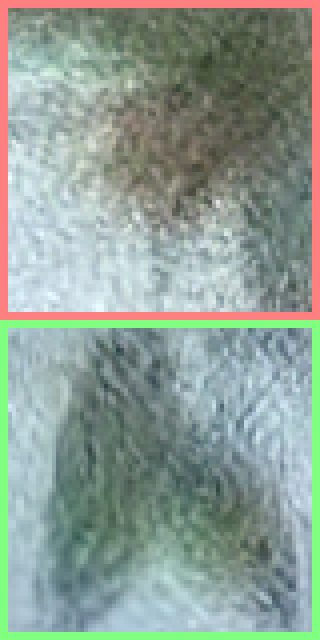}
		\includegraphics[width=\linewidth]{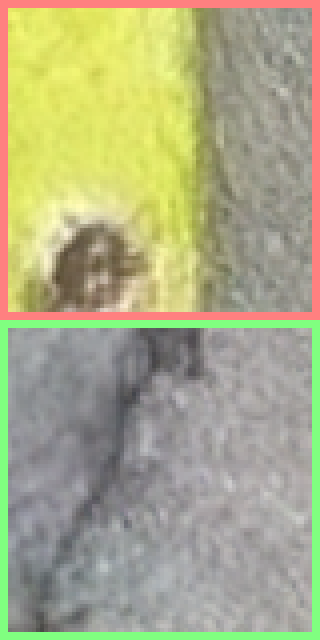}
	\end{minipage}%
	\begin{minipage}{\wid}%
		\includegraphics[width=\linewidth]{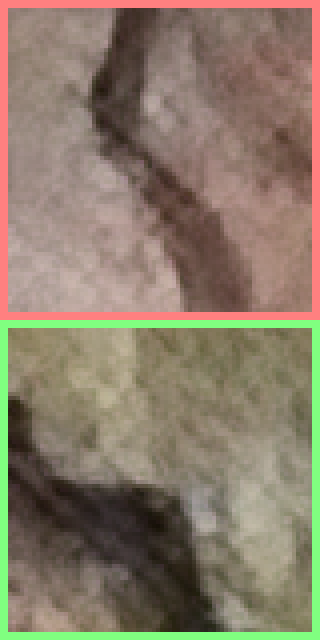}
		\includegraphics[width=\linewidth]{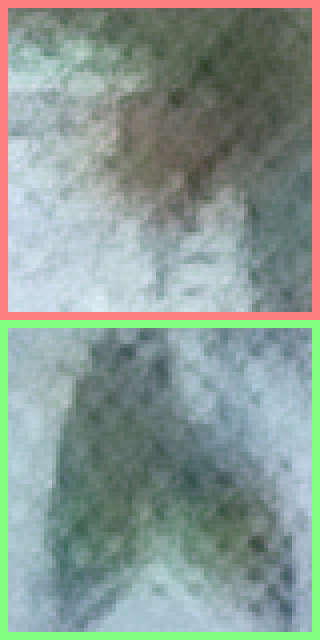}
		\includegraphics[width=\linewidth]{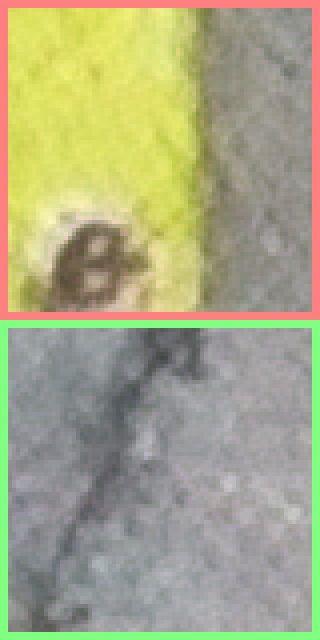}
	\end{minipage}%

	\newcommand{\sep}{~}
	\resizebox{\linewidth}{!}{
		\begin{tabular}{C{6cm}@{\sep}C{3cm}@{\sep}C{3cm}@{\sep}C{3cm}@{\sep}C{3cm}@{\sep}C{3cm}@{\sep}C{3cm}@{\sep}C{3cm}}
			Input & Input & \textbf{Ours} & ZSSR & EDSR & ESRGAN & ESRGAN FT
		\end{tabular}
	}
	\vspace{0mm}
	\caption{Qualitative comparison on real-world images from the DPED dataset (Section~4.4 in the paper). Our approach achieve superior perceptual quality. As this is the real-world dataset, the \textbf{ground-truth does not exist} for the super-resolution.}
	\label{fig:stateOfTheArt_dped}
\end{figure*}

\end{document}